\documentclass[11pt,pdftex, preprint]{elsarticle}
\usepackage[usenames,dvipsnames,svgnames,table]{xcolor}
\usepackage[colorlinks]{hyperref}
\AtBeginDocument{\hypersetup{linkcolor=blue, citecolor=blue}}
\usepackage{bm, bbm, bbding}
\usepackage{float}
\usepackage{amssymb,amsmath,amsfonts,mathrsfs,graphicx,kantlipsum,units,yfonts}
\usepackage{txfonts,dsfont}
\usepackage[toc,page,title,titletoc,header]{appendix}
\usepackage{setspace, framed, hyperref}
\usepackage[top=2.6cm, bottom=2.7cm, left=2.8cm, right=2.8cm]{geometry}
\usepackage{caption, subcaption}
\usepackage[novbox]{pdfsync}
\usepackage[T1]{fontenc}
\usepackage[font=footnotesize,labelfont=bf]{caption}

\usepackage[numbers]{natbib}

\makeatletter
\def\ps@pprintTitle{%
  \let\@oddhead\@empty
  \let\@evenhead\@empty
  \def\@oddfoot{\reset@font\hfil\thepage\hfil}
  \let\@evenfoot\@oddfoot
}
\makeatother

\journal{}
\begin{document}

\begin{frontmatter}

\title{Traversable wormholes and energy conditions in Lovelock-Brans-Dicke gravity}
\author{David Wenjie Tian\fnref{myfootnote}}
\address{Faculty of Science,  Memorial University, St. John's, Newfoundland, A1C 5S7, Canada}
\fntext[myfootnote]{Email address: wtian@mun.ca}

\begin{abstract}
Following the recent theory of Lovelock-Brans-Dicke gravity, we continue to investigate the conditions to support traversable wormholes by  the gravitational effects of spacetime parity and topology, which arise from the nonminimal couplings of a background scalar field to the Chern-Pontryagin density and the Gauss-Bonnet invariant. The flaring-out condition indicates that a Morris-Thorne-type wormhole can be maintained by violating the generalized null energy condition, and thus also breaking down the generalized weak, strong, and dominant energy conditions; meanwhile, analyses of the zero-tidal-force solution show that the standard null energy condition in general relativity can still be respected by the physical matter threading the wormhole. This way, the two topological effects have to dominate over the ordinary-matter source of gravity, and the scalar field is preferred to be noncanonical. By treating Brans-Dicke gravity as a reduced situation of Lovelock-Brans-Dicke gravity, we also examine the Brans-Dicke wormholes and energy conditions.\\

\noindent PACS numbers:  04.50.Kd, 04.20.Cv, 04.90.+e\\
\noindent Key words: traversable wormhole; Lovelock-Brans-Dicke gravity; Chern-Pontryagin density; Gauss-Bonnet invariant; flaring-out condition; generalized and standard energy conditions
\end{abstract}

\end{frontmatter}


\section{Introduction}

A wormhole is a fascinating passage as a shortcut connecting two distant regions in a spacetime or bridging two distinct universes. Pioneering investigations of wormholes can date back to the Einstein-Rosen bridge \cite{Einstein-Rosen bridge} in general relativity (GR), and earlier constructions of wormholes, such as those converted from the Kerr-Newman family of black holes, suffer from severe instability against small perturbations and immediate collapse of the throat after formation \cite{Matt Visser Wormholes book}.

Modern interest in wormholes are mainly based on the seminal work of Morris and Thorne on traversable Lorentzian wormholes \cite{Morris Thorne wormhole}, and the way to convert them into time machines \cite{Morris Thorne Yurtsever time machine}. Morris and Thorne firstly designed the metric with the desired structures of a traversable wormhole, and then recovered the matter fields through Einstein's equation. It turns out that the energy-momentum tensor has to violate the null energy condition, and thus it needs exotic matter to maintain the wormhole tunnel \cite{Morris Thorne wormhole}.
The standard energy conditions, however, are a cornerstone in many areas in GR, such as the classical black hole thermodynamics \cite{Hawking and Ellis, Poisson relativity toolkit}. Thus, much effort has been made to minimize the violation of the energy conditions and reduce the encounter of exotic matter at the throat (e.g. \cite{Morris Thorne wormhole, Rotating traversable wormholes, Matt Visser minimize NEC violation}).

The search for promising candidates of exotic matter is not an easy job, and only a small handful situations are recognized, such as the quantum Casimir effect and the semiclassical Hawking radiation, while all classical matter fields obey the standard energy conditions. With the development of precision cosmology and the discovery of cosmic acceleration, various models of dark energy with exotic equations of state have been proposed, which provide new possibilities to support wormholes, such as those supported by the cosmological constant \cite{Wormhole cosmological constant}, phantom- or quintom-type energy \cite{Wormhole Phantom, Wormhole Quintom}, generalized or modified Chaplygin gas \cite{Wormhole generalized Chaplygin gas, Wormhole modified Chaplygin gas}, and interacting dark sectors \cite{Wormhole interacting dark sectors}.

On the other hand, as an alternative to the mysterious dark energy, modified and alternative theories of relativistic gravity beyond GR have been greatly developed to explain the accelerated expansion of the Universe. The higher order terms or extra degrees of freedom in these theories yield antigravity effects, which overtake the gravitational attraction of ordinary matter at the cosmic scale. Lobo took Weyl conformal gravity as an example and suggested that modified gravities provide another possibility to support traversable wormholes \cite{Lobo Wormhole conformal}: it is the generalized energy conditions that are violated, while the standard energy conditions as in GR may remain valid. To date, this proposal has been applied to exact solutions of Morris-Thorne-type wormholes in various modified gravities, such as the metric $f(R)$ \cite{Lobo Wormhole fR}, nonminimal curvature-matter coupling \cite{Lobo Wormhole nonminimal curvature-matter coupling}, Brans-Dicke \cite{Lobo Wormhole Brans-Dicke}, modified teleparralel \cite{Lobo Wormhole teleparralel}, metric-Palatini hybrid $f(R)$ \cite{Lobo Wormhole hybrid metric-Palatini}, and Einstein-Gauss-Bonnet gravities \cite{Lobo Wormhole Einstein-Gauss-Bonnet}.

In this paper, we will look into traversable wormholes and the standard energy conditions in Lovelock-Brans-Dicke gravity \cite{Tian LBD gravity}, which takes into account the gravitational effects of spacetime parity and topology by the nonminimal couplings of a background scalar field to the Chern-Pontryagin density and the Gauss-Bonnet invariant.  This paper is organized as follows. We firstly review the gravity theory in Sec.~\ref{Sec Lovelock-Brans-Dicke gravity}, and derive its generalized energy conditions in Sec.~\ref{Sec LBD generalized energy conditions}. Then the conditions to support Morris-Thorne-type wormholes are investigated in Sec.~\ref{Sec Supporting wormholes in the LBD graity}, which are extensively examined by a zero-tidal-force solution in Sec.~\ref{Sec An exact solution}. Also, comparison with wormholes in Brans-Dicke gravity is studied in Sec.~\ref{Sec Wormholes in Brans-Dicke gravity}. Throughout this paper, we adopt the geometric conventions $\Gamma^\alpha_{\beta\gamma}=\Gamma^\alpha_{\;\;\,\beta\gamma}$,
$R^{\alpha}_{\;\;\beta\gamma\delta}=\partial_\gamma \Gamma^\alpha_{\delta\beta}-\partial_\delta \Gamma^\alpha_{\gamma\beta}\cdots$ and
$R_{\mu\nu}=R^\alpha_{\;\;\mu\alpha\nu}$ with the metric signature $(-,+++)$.


\section{Lovelock-Brans-Dicke gravity}\label{Sec Lovelock-Brans-Dicke gravity}

Recently we have discussed a new theory of alternative gravity which has been dubbed as Lovelock-Brans-Dicke (LBD) gravity \cite{Tian LBD gravity}. This theory is given by the action
\begin{equation}\label{Action LBD gravity}
\begin{split}
\mathcal{S}_{\text{LBD}}&=\int d^4x\sqrt{-g}\,\mathscr{L}_{\text{LBD}}+ \mathcal{S}_m\quad\mbox{with}\\
\mathscr{L}_{\text{LBD}}&=\frac{1}{16\pi}\Bigg[ \phi\left(R
+\frac{a}{\sqrt{-g}} {}^*RR + \hat{b}\mathcal{G}\right) -\frac{\omega_{\text{L}}}{\phi}\nabla_\alpha \phi
 \nabla^\alpha\phi-2V(\phi) \Bigg]\,,
\end{split}
\end{equation}
where $\phi=\phi(x^\alpha)$ is a background scalar field, $\{a,\hat{b}\}$ are dimensional coupling constants (note: $\hat{b}$ is hatted to be distinguished from $b=b(r)$ in Secs.~\ref{Sec Supporting wormholes in the LBD graity}, \ref{Sec An exact solution} and \ref{Sec Wormholes in Brans-Dicke gravity}, which is a standard denotation for the shape function in wormhole physics), $\omega_{\text{L}}$ denotes the dimensionless Lovelock parameter tuning the kinetics of $\phi(x^\alpha)$, $V(\phi) $ refers to a self-interaction potential, and as usual the matter action is given by
the matter Lagrangian density via $\mathcal{S}_m =\int d^4x \sqrt{-g}\,\mathscr{L}_m$. In Eq.(\ref{Action LBD gravity}), ${}^*RR$ and $\mathcal G$ denote the Chern-Pontryagin density
and the Gauss-Bonnet invariant, respectively,
\begin{equation}\label{Redefine CP}
\begin{split}
{}^*RR &\coloneqq
{}^{*}R_{\alpha\beta\gamma\delta} R^{\alpha\beta\gamma\delta}= \frac{1}{2} \epsilon_{\alpha\beta\mu\nu}
R^{\mu\nu}_{\;\;\;\;\,\gamma\delta}R^{\alpha\beta\gamma\delta}\,,\\
\mathcal{G} &\coloneqq R^2-4R_{\mu\nu}R^{\mu\nu}+R_{\mu\alpha\nu\beta}R^{\mu\alpha\nu\beta}\,,
\end{split}
\end{equation}
where ${}^{*}R_{\alpha\beta\gamma\delta}\coloneqq\frac{1}{2}\epsilon_{\alpha\beta\mu\nu} R^{\mu\nu}_{\;\;\;\;\gamma\delta}$ is the left dual of Riemann tensor, and $\epsilon_{\alpha\beta\mu\nu}$ represents the totally
antisymmetric Levi-Civita pseudotensor with $\epsilon_{0123}=\sqrt{-g}$
and $\epsilon^{0123}=1/\sqrt{-g}$. Note that unlike the other two curvature invariants $\left\{ R\right.$,
$\left. \mathcal{G}\right\}$, the term ${}^*RR$ in $\mathscr{L}_{\text{LBD}}$ is divided by $\sqrt{-g}$; this is
because ${}^*RR$ itself already serves as a covariant density for $\mathcal{S}_{\text{LBD}}$, as opposed to $\sqrt{-g}\, R $ and $ \sqrt{-g}\, \mathcal{G} $ therein.

$\mathcal{S}_{\text{LBD}}$ is inspired by the connection between GR and Brans-Dicke gravity, and proposed as the Brans-Dicke-type counterpart for the classic Lovelock action in Lovelock's theorem \cite{Book Lovelock Theorem}, i.e.
$
\mathcal{S}_{\text L}=\frac{1}{16\pi G}\int d^4x\sqrt{-g}\,\left(R-2\Lambda
+\frac{a}{\sqrt{-g}} {}^*RR + b\mathcal{G}\right)+ \mathcal{S}_m
$.
$\mathcal{S}_{\text L}$  is the most general action made up of algebraic curvature invariants
that yields second-order field equations in four dimensions, and limits the  field equation  to be Einstein's equation equipped with a cosmological constant $\Lambda$. The Chern-Pontryagin and the Gauss-Bonnet invariants in $\mathcal{S}_{\text L}$
do not influence the field equation, because ${}^*RR$ and $\sqrt{-g}\,\mathcal{G}$ are equal to the divergences
of their respective topological currents (see Ref.\cite{Tian LBD gravity} and the relevant references therein); instead, the nonminimally $\phi$--coupled
covariant densities $\phi{}^*RR$ and $\sqrt{-g}\,\phi\mathcal{G}$ in the LBD action Eq.(\ref{Action LBD gravity}) will have
nontrivial contributions to the field equation. Recall that for the two invariants ${}^*RR$ and $\mathcal{G}$, the former is related to the spacetime parity with $\int d^4x\,{}^*RR$
proportional to the instanton number of the spacetime, while the latter's integral  $ \frac{1}{32\pi^2}\int dx^4 \sqrt{-g}\,\mathcal{G}$
equates the Euler characteristic of the spacetime. Hence,  LBD gravity has taken into account the gravitational effects of the spacetime parity and the Euler topology.

The extremized variational derivative $\delta\mathcal{S}_{\text{LBD}}/\delta g^{\mu\nu}=0$ yields the gravitational field equation
\begin{equation}\label{LBD field equation}
\begin{split}
&\phi\left(R_{\mu\nu} -\frac{1}{2}  R  g_{\mu\nu}\right)
+ \left(g_{\mu\nu}\Box -\nabla_\mu \nabla_\nu \right) \phi  + a  H_{\mu\nu}^{\text{(CP)}} +\hat{b} H_{\mu\nu}^{\text{(GB)}}\\
&-\frac{\omega_{\text{L}}}{\phi}\left(\nabla_\mu \phi \nabla_\nu \phi-\frac{1}{2}g_{\mu\nu}\nabla_\alpha \phi
\nabla^\alpha\phi\right)+V(\phi)g_{\mu\nu}
=8\pi  T_{\mu\nu}^{\text{(m)}},
\end{split}
\end{equation}
where $H_{\mu\nu}^{\text{(CP)}}\coloneqq\frac{1}{\sqrt{-g}}\frac{\delta \left(\phi {}^*RR\right)}{\delta g^{\mu\nu}}$ collects the contributions
from the Chern-Pontryagin density with nonminimal coupling to $\phi(x^\alpha)$,
\begin{equation}\label{CP for field equation}
\sqrt{-g}\,H_{\mu\nu}^{\text{(CP)}}=\, 2\partial^\xi\phi\cdot
\left( \epsilon_{\xi\mu\alpha\beta}\nabla^\alpha R^\beta_{\;\;\nu} +
\epsilon_{\xi\nu\alpha\beta}\nabla^\alpha R^\beta_{\;\;\mu} \right) +2\partial_\alpha \partial_\beta  \phi \cdot \left({}^*R^{\alpha\;\;\,\beta}_{\;\;\,\mu\;\;\nu} +{}^*R^{\alpha\;\;\,\beta}_{\;\;\,\nu\;\;\mu}\right)\,,
\end{equation}
and $H_{\mu\nu}^{\text{(GB)}}\coloneqq\frac{1}{\sqrt{-g}}\frac{\delta(\sqrt{-g}\,\phi\mathcal{G})}{\delta g^{\mu\nu}}$ refers to the effect of extra degrees of freedom from the nonminimally $\phi-$coupled Gauss-Bonnet invariant,
\begin{equation} \label{GB for field equation}
\begin{split}
H_{\mu\nu}^{\text{(GB)}}=\,
&2R\left(g_{\mu\nu}\Box
-\nabla_\mu\!\nabla_\nu\right)\phi +4R_{\mu}^{\;\;\,\alpha}
\nabla_\alpha\!\nabla_{\nu}\phi +4R_{\nu}^{\;\;\,\alpha}\nabla_\alpha\!\nabla_{\mu} \phi \\
&-4R_{\mu\nu}\Box \phi -4g_{\mu\nu} \cdot R^{\alpha\beta}\nabla_\alpha\!\nabla_\beta \phi
+4R_{\alpha\mu \beta\nu}
\nabla^\beta \nabla^\alpha  \phi\,,
\end{split}
\end{equation}
with $\Box\coloneqq g^{\alpha\beta}\nabla_\alpha\nabla_\beta$ denoting the
covariant d'Alembertian. Compared to the field equations of the $f(R,\mathcal{G})$ and $f(R,\mathcal{G},\mathscr{L}_m)$ generalized Gauss-Bonnet gravities with
generic $\mathcal{G}-$dependence \cite{GaussBonnet modified gravity, Tian Gauss Bonnet paper}, we have removed the algebraic terms in $H_{\mu\nu}^{\text{(GB)}}$ by the Bach-Lanczos identity
$2 RR_{\mu\nu}-4 R_\mu^{\;\;\,\alpha}R_{\alpha\nu}
-4 R_{\alpha\mu\beta\nu}R^{\alpha\beta}
+2R_{\mu\alpha\beta\gamma}R_{\nu}^{\;\;\,\alpha\beta\gamma}
\equiv \frac{1}{2}\mathcal{G}g_{\mu\nu}$.
Immediately, the trace of the field equation (\ref{LBD field equation}) is found to be
\begin{equation}\label{Trace for field equation}
-\phi R
+\frac{\omega_{\text{L}}}{\phi}\nabla_\alpha \phi \nabla^\alpha\phi
+\left(3+2\hat{b}R\right) \Box \phi
-4 \hat{b} R^{\alpha\beta}\nabla_\alpha\!\nabla_\beta \phi +4V(\phi)=8\pi  T^{\text{(m)}},
\end{equation}
where $g^{\mu\nu} H_{\mu\nu}^{\text{(GB)}}=
2R \Box \phi - 4R^{\alpha \beta } \nabla_\alpha  \nabla_\beta \phi$, $T^{\text{(m)}}=g^{\mu\nu}T^{\text{(m)}}_{\mu\nu}$, and $H_{\mu\nu}^{\text{(CP)}}$ 
is always traceless.

On the other hand, for the scalar field $\phi(x^\alpha)$, the extremization $\delta\mathcal{S}_{\text{LBD}}/\delta\phi=0$ directly
leads to the \textit{kinematical} wave equation
\begin{equation}\label{kinematic wave equation for LBD}
2\omega_{\text{L}}  \Box\phi  =
-\left(R+\frac{a}{\sqrt{-g}}
 {}^*RR +\hat{b} \mathcal{G}\right)\phi+\frac{ \omega_{\text{L}}}{\phi}
\nabla_{\alpha}\phi \nabla^{\alpha} \phi+2V_\phi \phi\,,
\end{equation}
with $\Box\phi=g^{\alpha\beta}\nabla_\alpha\nabla_\beta \phi=\frac{1}{\sqrt{-g}}
\partial_\alpha\left(\sqrt{-g}\,g^{\alpha\beta}\partial_\beta \phi\right)$, and $V_\phi\coloneqq dV(\phi)/d\phi$.
Along with the trace equation (\ref{Trace for field equation}), it yields the \textit{dynamical} wave equation
\begin{equation}\label{dynamical wave equation for LBD}
\Big(2\omega_{\text{L}} +3+2\hat{b}R\Big)\Box\phi = -\left(
\frac{a}{\sqrt{-g}} {}^*RR +\hat{b} \mathcal{G}\right)\phi
 + 8\pi T^{\text{(m)}}
 + 4 \hat{b} R^{\alpha\beta}\nabla_\alpha\!\nabla_\beta \phi+2V_\phi \phi-4V(\phi)\,,
\end{equation}
which explicitly relates the propagation of $\phi(x^\alpha)$ to the trace $T^{\text{(m)}}$ of the matter tensor for the energy-momentum distribution.


In this paper, we will work out the conditions to support traversable wormholes in LBD gravity by the nontrivial gravitational effects of spacetime parity and topology due to the nonminimal $\phi$--couplings. To begin with, we firstly derive the generalized energy conditions for LBD gravity.


\section{Generalized LBD energy conditions}\label{Sec LBD generalized energy conditions}

In a region of a spacetime, for the expansion rate $\theta_{(\ell)}$ of a null congruence along its
null tangent vector field $\ell^\mu$, and the expansion rate $\theta_{(u)}$ of a
timelike congruence along its timelike tangent $u^\mu$,  $\theta_{(\ell)}$ and $\theta_{(u)}$ respectively satisfy the Raychaudhuri equations \cite{Poisson relativity toolkit}
\begin{align}
\ell^\mu \nabla_\mu \theta_{(\ell)} =\,\frac{d\theta_{(\ell)} }{d\lambda}&\:=
\:\kappa_{(\ell)} \theta_{(\ell)} - \frac{1}{2}\,\theta^2_{(\ell)}
-\sigma_{\mu\nu}^{(\ell)} \sigma^{\mu\nu}_{(\ell)} +\omega_{\mu\nu}^{(\ell)}
\omega^{\mu\nu}_{(\ell)} -R_{\mu\nu}\ell^\mu \ell^\nu\,,\\
u^\mu \nabla_\mu \theta_{(u)}=\frac{d\theta_{(u)}}{d\tau}&\,=\,
\kappa_{(u)}  \theta_{(u)}- \frac{1}{3}\,\theta^2_{(u)}-\sigma_{\mu\nu}^{(u)}
\sigma^{\mu\nu}_{(u)}
+\omega_{\mu\nu}^{(u)}
\omega^{\mu\nu}_{(u)}-R_{\mu\nu}u^\mu u^\nu\,.
\end{align}
The inaffinity coefficients are zero $\kappa_{(\ell)}=0=\kappa_{(u)}$ under
affine parameterizations, the twist vanishes $\omega_{\mu\nu}\omega^{\mu\nu}=0$ for
hypersurface-orthogonal foliations, and being spatial tensors $\left(\sigma_{\mu\nu}^{(\ell)}\ell^\mu=0
=\sigma_{\mu\nu}^{(u)}u^\mu\right)$ the shears
always satisfy $\sigma_{\mu\nu}\sigma^{\mu\nu}\geq 0$.
Thus, to guarantee $d\theta_{(\ell)}/
d\lambda\leq 0$ and $d\theta_{(u)}
/d\tau\leq 0$ under all circumstances -- even in the occasions $\theta_{(\ell)}=0=\theta_{(u)}$, so that the congruences
focus and gravity is always an attractive force, the following geometric
nonnegativity conditions are expected to hold:
\begin{equation}\label{GECs Geometric}
R_{\mu\nu}\ell^\mu 
\ell^\nu \,\geq\,0 \qquad,\qquad
R_{\mu\nu}u^\mu u^\nu 
\,\geq\,0 \,.
\end{equation}

Note that although this is the most popular approach to derive Eq.(\ref{GECs Geometric}) for its straightforwardness and simplicity, it is not perfect. In general $\theta_{(\ell)}$ and $\theta_{(u)}$ are nonzero and one could only obtain $\frac{1}{2}\theta^2_{(\ell)} 
+R_{\mu\nu}\ell^\mu \ell^\nu\geq 0$ and $\frac{1}{3}\theta^2_{(u)}
+R_{\mu\nu}u^\mu u^\nu\geq 0$. Thus, it is only safe to say that Eq.(\ref{GECs Geometric}) provides the sufficient rather than necessary conditions to ensure
$d\theta_{(\ell)}/d\lambda\leq 0$ and  $d\theta_{(u)}/d\tau\leq 0$. Fortunately, this imperfectness is not a disaster and does not negate the conditions in Eq.(\ref{GECs Geometric}); for example, one can refer to Ref. \cite{Derivation Null EC} for a rigorous derivation of the first inequality in  Eq.(\ref{GECs Geometric}) from the Virasoro
constraint in the worldsheet string theory.

On the other hand, consider generic relativistic gravities with the Lagrangian density $\mathscr{L}_{\text{total}}=\frac{1}{16\pi G}\mathscr{L}_G $ $(R,R_{\mu\nu}R^{\mu\nu},\mathcal{R}_{\,i}\,,\cdots,
\vartheta, \nabla_\mu\vartheta\nabla^\mu\vartheta\big)+\mathscr{L}_m $, where $\mathcal{R}_i=\mathcal{R}_i\,\big(g_{\alpha\beta}\,,R_{\mu\alpha\nu\beta}\,,
\nabla_\gamma R_{\mu\alpha\nu\beta}\,,\ldots\big)$ refers to a generic curvature invariant beyond the Ricci scalar, and $\vartheta$ denotes a scalarial
extra degree of freedom
unabsorbed by $\mathscr{L}_m$. The field equation reads
\begin{equation}\label{Field Equation in GR Form 0}
\mathcal{H}_{\mu\nu}=8\pi G T_{\mu\nu}^{\text{(m)}}\;\quad
\mbox{with}\;\quad
\mathcal{H}_{\mu\nu}\,\coloneqq\, \frac{1}{\sqrt{-g}}
\frac{\delta\, \Big(\!\!\sqrt{-g}\,\mathscr{L}_G \Big)}
{\delta g^{\mu\nu}}\,,
\end{equation}
where total-derivative terms should be removed in the derivation of $\mathcal{H}_{\mu\nu}$. In the spirit of  reconstructing an effective dark energy,
Eq.(\ref{Field Equation in GR Form 0}) can be intrinsically recast into a compact GR form by
isolating the Ricci tensor $R_{\mu\nu}$ out of $\mathcal{H}_{\mu\nu}$:
\begin{equation}\label{MG Field Equation in GR Form}
R_{\mu\nu} -\frac{1}{2}Rg_{\mu\nu} = 8\pi G_{\text{eff}}
T_{\mu\nu}^{\text{(eff)}}\;\quad\text{with}\;\quad
\mathcal{H}_{\mu\nu}=\frac{G}{G_{\text{eff}}}
G_{\mu\nu}-8\pi G T_{\mu\nu}^{\text{(MG)}}\,,
\end{equation}
where $G_{\text{eff}}$ denotes the effective gravitational coupling strength, and it is recognized
from the coefficient of the matter tensor $T_{\mu\nu}^{\text{(m)}}$.
$T_{\mu\nu}^{\text{(eff)}}$ refers to the total effective energy-momentum tensor,
and $T_{\mu\nu}^{\text{(MG)}} = T_{\mu\nu}^{\text{(eff)}}-T_{\mu\nu}^{\text{(m)}}$, with
$T_{\mu\nu}^{\text{(MG)}}$ collecting all the
modified-gravity nonlinear and higher-order effects. Thus, all terms beyond GR have been packed into $T_{\mu\nu}^{\text{(MG)}}$ and $G_{\text{eff}}$.

Following Eq.(\ref{MG Field Equation in GR Form})
along with its trace equation $R =-8\pi G_{\text{eff}}   T^{\text{(eff)}}$
and the equivalent form $R_{\mu\nu} = 8\pi G_{\text{eff}}  \left(T_{\mu\nu}^{\text{(eff)}}-\frac{1}{2} g_{\mu\nu}T^{\text{(eff)}} \right)$,
the geometric nonnegativity conditions in Eq.(\ref{GECs Geometric}) can be translated
into the generalized null and strong energy conditions (GNEC and GSEC for short)
\begin{equation}\label{Null and Stong GECs}
G_{\text{eff}}T_{\mu\nu}^{\text{(eff)}}\,\ell^\mu \ell^\nu \,\geq\,0\quad (\text{GNEC})\quad,\quad
G_{\text{eff}}  \left(T_{\mu\nu}^{\text{(eff)}}u^\mu u^\nu+\frac{1}{2} T^{\text{(eff)}} \right)\,\geq\,0
\quad (\text{GSEC})\,,
\end{equation}
where $\ell^\mu \ell_\mu=0$ for the GNEC, and $u_\mu u^\mu=-1$  in the GSEC for compatibility with the metric signature $(-,+++)$. We further supplement Eq.(\ref{Null and Stong GECs}) by the generalized weak energy condition 
\begin{equation}\label{Weak GEC}
G_{\text{eff}}T_{\mu\nu}^{\text{(eff)}} u^\mu u^\nu \,\geq\,0\quad (\text{GWEC})\,,
\end{equation}
and the generalized dominant energy condition (GDEC) that
$G_{\text{eff}}T_{\mu\nu}^{\text{(eff)}} u^\mu u^\nu\geq 0$ with $G_{\text{eff}}T_{\mu\nu}^{\text{(eff)}} u^\mu $ being a causal vector.

Note that for the common pattern of the field equations in modified gravities, we have chosen to adopt Eq.(\ref{MG Field Equation in GR Form}) rather than $R_{\mu\nu}-\frac{1}{2}Rg_{\mu\nu}=8\pi G\widehat{T}_{\mu\nu}^{\text{(eff)}}$, where $G$ is Newton's constant. That is to say, we do not absorb $G_{\text{eff}}$ into $T_{\mu\nu}^{\text{(eff)}}$ so that  $G_{\text{eff}} T_{\mu\nu}^{\text{(eff)}}=G \widehat{T}_{\mu\nu}^{\text{(eff)}}$; as a consequence, $G_{\text{eff}}$ shows up in the generalized energy conditions as well. This is because the effective matter-gravity coupling strength $G_{\text{eff}}$ plays important roles in many physics problems, such as the Wald entropy of black-hole horizons \cite{Wald entropy BH horizons}
and the cosmological gravitational thermodynamics (e.g.\cite{Tian gravitational thermodynamics}), although the meanings and applications of $G_{\text{eff}}$ have not been fully understood (say the relations between $G_{\text{eff}}$ and the weak, Einstein, and strong equivalence principles).

$G_{\text{eff}}$ and $T_{\mu\nu}^{\text{(eff)}}$ vary among different theories of modified gravity, which concretize Eqs.(\ref{Null and Stong GECs}) and (\ref{Weak GEC}) into different sets of generalized energy conditions. For LBD gravity summarized in Sec.~\ref{Sec Lovelock-Brans-Dicke gravity}, we have
\begin{equation}
\begin{split}
G_{\text{eff}}=\phi^{-1} \quad\text{and}\quad T_{\mu\nu}^{\text{(eff)}}= T_{\mu\nu}^{\text{(m)}}+T_{\mu\nu}^{(\phi)}
+T_{\mu\nu}^{\text{(CP)}}+T_{\mu\nu}^{\text{(GB)}},
\end{split}
\end{equation}
with the components of $T_{\mu\nu}^{\text{(eff)}}$ given by
\begin{equation}\label{Effective Tmunu phi CP GB}
\begin{split}
&8\pi T_{\mu\nu}^{\text{(CP)}}=-a  H_{\mu\nu}^{\text{(CP)}}\;\;\;,\;\;\;
8\pi T_{\mu\nu}^{\text{(GB)}}=-\hat{b} H_{\mu\nu}^{\text{(GB)}}\,,\\
8\pi T_{\mu\nu}^{(\phi)}&= \left(\nabla_\mu \nabla_\nu- g_{\mu\nu}\Box\right)\phi
+\frac{\omega_{\text{L}}}{\phi}
\left(\nabla_\mu \phi \nabla_\nu \phi-\frac{1}{2}g_{\mu\nu}\nabla_\alpha \phi
\nabla^\alpha\phi\right)-Vg_{\mu\nu}\,.
\end{split}
\end{equation}
Hence, for LBD gravity, the GNEC, GWEC and GSEC are respectively
\begin{equation}\label{LBD Null GEC}
\begin{split}
\phi^{-1}\ell^\mu \ell^\nu\Bigg(
8\pi T_{\mu\nu}^{\text{(m)}}
+\nabla_\mu \nabla_\nu\phi
+\frac{\omega_{\text{L}}}{\phi}\nabla_\mu \phi \nabla_\nu \phi
-a H_{\mu\nu}^{\text{(CP)}}
-\hat{b}H_{\mu\nu}^{\text{(GB)}}
\Bigg)\,\geq\,0\,,
\end{split}
\end{equation}
\begin{equation}\label{LBD Weak GEC}
\begin{split}
\hspace{-2.8mm}\phi^{-1}u^\mu u^\nu\left(
8\pi T_{\mu\nu}^{\text{(m)}}
+\nabla_\mu \nabla_\nu\phi
+\frac{\omega_{\text{L}}}{\phi} \nabla_\mu \phi \nabla_\nu \phi
- a H_{\mu\nu}^{\text{(CP)}}
-\hat{b}H_{\mu\nu}^{\text{(GB)}}
\right)+\phi^{-1}\left( \Box\phi + \frac{\omega_{\text{L}}}{2\phi }\nabla_\alpha \phi \nabla^\alpha\phi+ V \right)\geq 0,
\end{split}
\end{equation}
\begin{flalign}\label{LBD strong GEC}
\text{and}\qquad\qquad\qquad\quad
\phi^{-1}u^\mu u^\nu&\left(
8\pi T_{\mu\nu}^{\text{(m)}}
+\nabla_\mu \nabla_\nu\phi
+\frac{\omega_{\text{L}}}{\phi} \nabla_\mu \phi \nabla_\nu \phi
- a H_{\mu\nu}^{\text{(CP)}}
-\hat{b}H_{\mu\nu}^{\text{(GB)}}
\right)&&\nonumber\\
+\frac{1}{2}\phi^{-1}&\bigg(8\pi T^{\text{(m)}}+
4 \hat{b} R^{\alpha\beta}\nabla_\alpha\!\nabla_\beta \phi
-(1+2\hat{b}R) \Box \phi -2V\bigg)\,\geq\,0\,,&&
\end{flalign}
while the GDEC can be concretized in the same way.
Among all generalized energy conditions, Eq.(\ref{LBD Null GEC}) clearly shows that the GNEC is not influenced by the background potential $V=V(\phi)$ of the scalar field.

Particularly, LBD gravity reduces to become GR for the situation $\phi(x^\alpha)\equiv G^{-1}=\text{constant}$ and $V(\phi)=0$, as ${}^*RR $ and $\sqrt{-g}\mathcal{G}$ in Lovelock's action $\mathcal{S}_{\text L}$ do not affect the field equation. Then Eqs.(\ref{Null and Stong GECs}) and (\ref{Weak GEC}) reduce to become the standard energy conditions for classical matter fields \cite{Hawking and Ellis}:
\begin{equation}\label{GR Null and Stong ECs}
T_{\mu\nu}^{\text{(m)}}\,\ell^\mu \ell^\nu \,\geq\,0\quad (\text{NEC})\quad,\quad
T_{\mu\nu}^{\text{(m)}}\,u^\mu u^\nu \,\geq\,0\quad (\text{WEC})
\quad,\quad
T_{\mu\nu}^{\text{(m)}}u^\mu u^\nu\,\geq\,\frac{1}{2}\, T^{\text{(eff)}}u_\mu u^\mu\quad (\text{SEC})\,.
\end{equation}


\section{Conditions to support wormholes in LBD graity}\label{Sec Supporting wormholes in the LBD graity}

\subsection{Generic conditions supporting static, spherically symmetric wormholes}\label{Subsec Generic conditions supporting static, spherically symmetric wormholes}
\vspace{2mm}

It has been nearly three decades since the classical work of Morris and Thorne, and nowadays the Morris-Thorne metric for static spherically symmetric wormholes is still the most useful and popular ansatz to study traversable wormholes. The metric reads \cite{Morris Thorne wormhole}
\begin{equation}\label{Morris Thorne wormhole ansatz}
ds^2=-e^{2\Phi(r)} dt^2+  \left(1-\frac{b(r)}{r}\right)^{-1}dr^2
+r^2\left(d\theta^2+\sin^2\theta  d\varphi^2\right)\,,
\end{equation}
where $\Phi(r)$ and $b(r)$ are the redshift and the shape functions, respectively, and the radial coordinate $r\geq r_0$ ranges from a minimum value $r_0$ at the wormhole throat to infinity. $\Phi(r)$ is related to the gravitational redshift of an infalling body, and  it must be finite everywhere to avoid the behavior $e^{2\Phi(r)}\to 0$ and consequently the existence of an event horizon. $b(r)$ determines the shape of the 2-slice $\{t=\mbox{constant}, \theta=\pi/2\}$ in the embedding diagram; it satisfies $b(r)<r$ to keep the wormhole Lorentzian, $b(r_0)=r_0$ at the throat, and $b(r)/r\to 0$ at $r\to\infty$ if asymptotically flat. Moreover, the embedding of the 2-slice $ds^2= \left(1-\frac{b(r)}{r}\right)^{-1}dr^2+r^2 d\varphi^2$ yields the
geometrical ``flaring-out condition'' $(b- b' r)/b^2>0$, which  reduces to become $b'(r_0)<1$ at the throat $r=r_0$ with  $b(r_0)=r_0$ \cite{Morris Thorne wormhole}. Here and hereafter the prime denotes the derivative with respect to the radial coordinate $r$.

Following the metric Eq.(\ref{Morris Thorne wormhole ansatz}), in the null tetrad  adapted to the spherical symmetry and the null radial congruence,
\begin{equation}
\begin{split}
\vspace{-2mm}l^\mu=\left(e^{-\Phi(r)},\sqrt{1-\frac{b(r)}{r}}\,,0,0 \right)
\;,\;\; &n^\mu=\frac{1}{2}\,\left(-e^{-\Phi(r)},\sqrt{1-\frac{b(r)}{r}}\,,0,0 \right)
\; ,\;\; m^\mu=\frac{1}{\sqrt{2}\,r}\,\Big(0,0,1,\frac{i}{\sin\!\theta}\Big),
\end{split}
\end{equation}
one could find the outgoing expansion rate $\theta_{(l)}$ and the ingoing expansion rate $\theta_{(n)}$ to be
\begin{equation}\label{Wormhole expansion rates}
\theta_{(l)} = -\left(\rho_{\text{NP}}+\bar{\rho}_{\text{NP}}\right) = \frac{2}{r}\sqrt{1-\frac{b(r)}{r}}
\;\;\quad,\;\;\quad
\theta_{(n)}=\mu_{\text{NP}}+\bar{\mu}_{\text{NP}}=\frac{1}{r}\sqrt{1-\frac{b(r)}{r}},
\end{equation}
where $\rho_{\text{NP}} \coloneqq -m^\mu \bar{m}^\nu \nabla_\nu \ell_\mu$
and $\mu_{\text{NP}} \coloneqq \bar{m}^\mu m^\nu\nabla_\nu n_\mu$ are two Newman-Penrose spin coefficients.
Thus the metric ansatz Eq.(\ref{Morris Thorne wormhole ansatz}) guarantees that the spacetime is everywhere antitrapped as $\theta_{(l)}=2\theta_{(n)}>0$, which is a characteristic property of traversable wormholes and resembles white holes \cite{Hochberg Visser}. Also,  Eq.(\ref{Wormhole expansion rates}) shows that the expansion rates are independent of the redshift function $\Phi(r)$, and the spacetime is free of apparent horizons for $r>r_0$.

Since the outward-flaring constraint $(b- b' r)/b^2>0$ solely comes from the  embedding geometry, it is independent of and applicable to all gravity theories. In GR through Einstein's equation, this condition implies that all infalling observers threading a Morris-Thorne wormhole will experience the violation of the standard null energy condition $T_{\mu\nu}^{\text{(m)}}\ell^\mu \ell^\nu\geq0$ \cite{Morris Thorne wormhole, Morris Thorne Yurtsever time machine}. Similarly, according to the GR form of the field equation (\ref{MG Field Equation in GR Form}),
the flaring-out condition implies that wormholes in LBD gravity are supported by  the breakdown of the LBD generalized energy conditions as in Eqs.(\ref{LBD Null GEC}) and (\ref{LBD Weak GEC}). On the other hand,
in principle it may still be possible to preserve the standard energy conditions in Eq.(\ref{GR Null and Stong ECs}). Thus, to fulfill the constraint $(b- b' r)/b^2>0$ in LBD gravity, a possible way to violate the GNEC while keeping the standard NEC, i.e. $\phi^{-1}T_{\mu\nu}^{\text{(eff)}} \ell^\mu \ell^\nu <0$ and $T_{\mu\nu}^{\text{(m)}} \ell^\mu \ell^\nu \geq 0$, can be
\begin{equation}\label{Wormhole GEC Null}
\begin{split}
0\,\leq\,8\pi \ell^\mu \ell^\nu T_{\mu\nu}^{\text{(m)}}
\,\leq\,\ell^\mu \ell^\nu\Bigg(
a  H_{\mu\nu}^{\text{(CP)}}
+\hat{b}H_{\mu\nu}^{\text{(GB)}}
-\nabla_\mu \nabla_\nu\phi
-\frac{\omega_{\text{L}}}{\phi}\nabla_\mu \phi \nabla_\nu \phi
\Bigg)\,.
\end{split}
\end{equation}
As another example, violation of the GWEC and preservation of the WEC,  i.e. $\phi^{-1}T_{\mu\nu}^{\text{(eff)}} u^\mu u^\nu < 0$ and $T_{\mu\nu}^{\text{(m)}} u^\mu u^\nu \geq 0$,  can be realized if
\begin{equation}\label{Wormhole Weak GEC}
\begin{split}
0<8\pi T_{\mu\nu}^{\text{(m)}}u^\mu u^\nu<
u^\mu u^\nu\left(a H_{\mu\nu}^{\text{(CP)}}
+\hat{b}H_{\mu\nu}^{\text{(GB)}}
-\nabla_\mu \nabla_\nu\phi
-\frac{\omega_{\text{L}}}{\phi} \nabla_\mu \phi \nabla_\nu \phi
\right)-\left(\frac{\omega_{\text{L}}}{2\phi }\nabla_\alpha \phi \nabla^\alpha\phi+ \Box\phi + V\right).
\end{split}
\end{equation}
Eqs.(\ref{Wormhole GEC Null}) and (\ref{Wormhole Weak GEC}) indicate that $H_{\mu\nu}^{\text{(CP)}}$ and $H_{\mu\nu}^{\text{(GB)}}$, which represent the effects of the spacetime parity and topology, jointly with $T_{\mu\nu}^{(\phi)}$ should dominate over the material source of gravity. Also, a noncanonical scalar field ($\omega_{\text{L}}<0$) is preferred than a canonical one ($\omega_{\text{L}}>0$) to help support the wormhole.

Note that in Eqs.(\ref{Wormhole GEC Null}) and (\ref{Wormhole Weak GEC})  we have assumed $\phi^{-1}=G_{\text{eff}}>0$. This is inspired by the fact in $f(R)$ gravity that the effective coupling strength $G_{\text{eff}}=df(R)/dR\eqqcolon f_R$ has to satisfy $f_R>0$ to guarantee that in the particle content via the spin projectors, the graviton itself and the induced scalar particle are not ghosts \cite{Geff for fR positive}.  Similarly, in scalar-tensor gravity $\mathscr{L}=\frac{1}{16\pi G}\left[f(\phi)R-h(\phi)\nabla_\alpha\phi\nabla^\alpha\phi-2U(\phi) \right]+\mathscr{L}_m$ in the Jordan frame,
$G_{\text{eff}}=f(\phi)^{-1}$ should also be positive definite so that the graviton is not a ghost \cite{Positive Geff scalar-tensor}. More generally, for modified gravities of the field equation (\ref{MG Field Equation in GR Form}), an assumption $G_{\text{eff}}>0$ can not only simplify the generalized energy conditions Eqs.(\ref{Null and Stong GECs}) and (\ref{Weak GEC}), but also help reduce the violation of these conditions.


In Sec.~\ref{Sec An exact solution}, we will demonstrate by a zero-tidal-force solution that Eq.(\ref{Wormhole GEC Null}) can really be satisfied while Eq.(\ref{Wormhole Weak GEC}) is partially falsified for the same numerical setups. To facilitate the discussion, we further concretize the tensorial inequalities Eqs.(\ref{Wormhole GEC Null}) and (\ref{Wormhole Weak GEC})  into an anisotropic perfect fluid form.


\subsection{Supporting conditions in anisotropic fluid scenario}\label{Subsec Supporting conditions in anisotropic fluid scenario}

\vspace{2mm}

In accordance with the nonzero and unequal components of the Einstein tensor $G^{\mu}_{\;\;\nu}$, one can assume an anisotropic perfect-fluid form $T^{\mu}_{\;\;\nu}=\text{diag}\left[-\rho(r),P^r(r),P^T(r),P^T(r)\right]$ for $T^{\mu\,\text{(eff)}}_{\;\;\nu}$ and each of its components. Here $T^{\mu}_{\;\;\nu}$ is adapted to the metric signature $(-,+++)$, with $\rho$ standing for the energy density, $P^r$ for the radial pressure, and $P^T$ for the transverse pressure orthogonal to the radial direction.
In wormhole physics, it is $P^r$ that helps to open and maintain the wormhole tunnel, so in the context below we will be more concentrative on $P^r$ rather than $P^T$.
Then the generalized energy conditions in Sec.~\ref{Sec LBD generalized energy conditions} imply $G_{\text{eff}}(\rho_{\text{eff}}+P^r_{\text{eff}})\geq 0$ for the GNEC, $G_{\text{eff}}\rho_{\text{eff}}\geq0$ and $G_{\text{eff}}(\rho_{\text{eff}}+P^r_{\text{eff}})\geq 0$ for the GWEC,   $G_{\text{eff}}(\rho_{\text{eff}}+P^r_{\text{eff}}+2P^T_{\text{eff}})\geq0$ and  $G_{\text{eff}}(\rho_{\text{eff}}+P^r_{\text{eff}})\geq 0$ for the GSEC, as well as $G_{\text{eff}}\rho_{\text{eff}}\geq0$ and $G_{\text{eff}} \rho_{\text{eff}}\geq \left|G_{\text{eff}} P^r_{\text{eff}}\right|$ for the GDEC, with $G_{\text{eff}}$ removable when $G_{\text{eff}}>0$.


In fact, the perfect-fluid form of $T^{\mu}_{\;\;\nu}$ clearly shows that violation of the null energy condition -- that is to say, giving up the dominance of the energy density over the pressure, will imply the simultaneous violations of the weak, strong, and dominant energy conditions. This chain of violation happens for both the standard and the generalized energy conditions, and in this sense, it is sufficient to consider the violation of the null energy condition. According to the GNEC in LBD gravity, it requires $\phi^{-1}(\rho_{\text{eff}}+P^r_{\text{eff}})< 0$ to make wormholes flare outward, with $\rho_{\text{eff}}=\rho_m+\rho_\phi+a\rho_{\text{CP}}+\hat{b}\rho_{\text{GB}}$ and $P^r_{\text{eff}}=P_m^r+P_\phi^r+aP_{\text{CP}}^r+\hat{b}P_{\text{GB}}^r$;
under the Morris-Thorne metric, we have
$\rho_{\text{CP}}=0=P_{\text{CP}}^r$,
\begin{equation}\label{Generic rho P 1}
8\pi\rho_\phi=
\left(1-\frac{b}{r}\right) \left(\Phi'\phi'+\phi''+\frac{2\phi'}{r}+\frac{\omega_{\text L}}{2} \frac{\phi'^2}{\phi} \right)
+\frac{\phi'}{2r^2}\left(b- b'r\right) +V \,,
\end{equation}
\begin{equation}\label{Generic rho P 2}
8\pi P^r_{\phi}=
\left( 1 -\frac{b }{ r}\right)\left(\phi''+\omega_{\text L}\frac{ \phi'^2 }{\phi } \right)-8\pi\rho_\phi \,,
\end{equation}
\begin{eqnarray}
8\pi\rho_{\text{GB}} =\frac{1}{r^5}\,\bigg[\!\!\! & -&\!\!\!4br\phi''(b-r) - 2\phi' (2 r- 3b)(b-b'r) +4\Phi'^2\phi' r^2 \left( bb' r + 2 r^2 - b' r^2+  b^2 - 3b r      \right)\nonumber \\
&+&\!\!\!2\Phi''\phi' r^2 \left( bb' r -  b' r^2 - b^2  + b r     \right)
 +4\Phi'\Phi''\phi' r^3 \left( b- r\right)^2+\Phi'^3 r^3 \left( r^2 -8b r +4 b^2\right)\nonumber  \\
&+&\!\!\!\Phi'\phi' r \left( 8r^2 -16b r +4 b' r^2  + b'^2 r^2 -6b b' r +9b^2 \right)
 \bigg]\,,\quad\mbox{and}\label{Generic rho P 3}
\end{eqnarray}
\begin{equation}\label{Generic rho P 4}
\begin{split}
8\pi P^r_{\text{GB}}= \frac{2}{r^5}\,\bigg[
&\phi'\left(b - b' r\right)\left(3b-4 r + b' r\right)
+2\phi''br(r-b)+4\Phi'\phi' r\left( \Phi' r  + 2 \right)\left(b- r \right)^2
\bigg]-8\pi \rho_{\text{GB}}\,.
 \end{split}
\end{equation}




\section{Zero-tidal-force solution}\label{Sec An exact solution}

In this section we will continue to work out an exact solution of Morris-Thorne wormholes in LBD gravity, so as to better analyze the flaring-out condition for the wormhole throat, and examine the states of the generalized and standard energy conditions.

There are two functions to be specified in the Morris-Thorne metric Eq.(\ref{Morris Thorne wormhole ansatz}). To be more concentrative on the wormhole throat and the embedding geometry, we will consider a zero redshift function $\Phi(r)=0$ or $e^{2\Phi(r)}=1$, which corresponds to vanishing tidal force and stationary observers \cite{Morris Thorne wormhole}. In this situation, the LBD curvature invariants and the Einstein tensor read
\begin{equation}\label{curvature invariants and Einstein tensor}
\begin{split}
R=\frac{2b'}{r^2}\,,\quad{}^*RR=0=\mathcal{G}\,,\quad G^\mu_{\;\;\nu}=r^{-3}\cdot \text{diag}\Big[ -b' r ,\; -b  ,\; b-b'r  ,\; b-b'r\Big]\,,
\end{split}
\end{equation}
and thus the componential field equations $G^\mu_{\;\;\nu}=8\pi \phi^{-1}T^{\mu\,\text{(eff)}}_{\;\;\nu}$ directly illustrate the influences of the flaring-out condition $(b-b'r)/b^2>0$ to $T^{\mu\,\text{(eff)}}_{\;\;\nu}=
\text{diag}\left[-\rho_{\text{eff}},P^r_{\text{eff}},P^T_{\text{eff}},
P^T_{\text{eff}}\right]$.

To simplify the dynamical wave equation (\ref{dynamical wave equation for LBD}), we assume the potential $V(\phi)$ to satisfy the condition $V_\phi \phi=2V$, which integrates to yield
\begin{equation}
\begin{split}
V(\phi)=V_0\phi^2\,,
\end{split}
\end{equation}
where $V_0$ is an integration constant. Moreover, we adopt the following power-law ansatz for the static and spherically symmetric scalar field,
\begin{equation}\label{Homogeneous scalar field ansatz}
\begin{split}
\phi(r)=\phi_0 \left(\frac{r_0}{r}\right)^A\,,
\end{split}
\end{equation}
where $\phi_0$ and the power index $A$ are constants, and $r_0$ is the throat radius $r_0=\min(r)$.


With these setups, the kinematical wave equation (\ref{kinematic wave equation for LBD}) leads to
\begin{equation}
\begin{split}
\left(2+\omega_{\text L} A \right)r b'
-\omega_{\text L}A(A-1)b
+\omega_{\text L}A(A-2)r
-4V_0\phi_0 \left(\frac{r_0}{r}\right)^A r^3=0\,.
\end{split}
\end{equation}
Solving this equation for $b(r)$ with the boundary condition $b(r=r_0)=r_0$,
we obtain  the shape function
\begin{equation}\label{Exact solution br}
\begin{split}
b(r)=\frac{2 V_0\phi_0 r_0^3
\left[
\left(\frac{r_0} {r }\right)^{\frac{\omega_{\text L}A(1-A) }{\omega_{\text L}A+2}} -\left(\frac{r_0}{r}\right)^{A-3}  \right]}{\omega_{\text L}A^2-2\omega_{\text L}A+A-3}
+
\frac{\omega_{\text L}A \left(A-2 \right)r-2r_0\left(\frac{r_0} {r }\right)^{\frac{\omega_{\text L}A(1-A) }{\omega_{\text L}A+2}}}
{\omega_{\text L}A^2-2\omega_{\text L}A-2 } \,,
\end{split}
\end{equation}
and thus
\begin{equation}\label{Exact solution b-br'}
\begin{split}
b-b'r=&\frac{2 V_0\phi_0 r_0^3
\left[\left(1+\frac{\omega_{\text L}A(1-A) }{\omega_{\text L}A+2}\right)\left(\frac{r_0} {r }\right)^{\frac{\omega_{\text L}A(1-A)}{\omega_{\text L}A+2}}
-(A-2)\left(\frac{r_0}{r}\right)^{A-3} \right]}{\omega_{\text L}A^2-2\omega_{\text L}A+A-3}
+\frac{2r_0}
{\omega_{\text L}A+2}\left(\frac{r_0} {r }\right)^{\frac{\omega_{\text L}A(1-A) }{\omega_{\text L}A+2}}\,.
\end{split}
\end{equation}
At the throat $r=r_0$, the flaring-out constraint $(b-b'r)/b^2$ is evaluated as
\begin{equation}\label{Throat constraint on parameters}
\begin{split}
b'(r_0)=\frac{4V_0\phi_0 r_0^2+\omega_{\text L}A}{\omega_{\text L}A+2}<1\,.
\end{split}
\end{equation}

There are five parameters in $b(r)$, among which  $\left\{r_0, \phi_0, A, V_0 \right\}$ attribute to our solution ansatz for the homogeneous scalar field $\phi(r)$ and the potential $V(\phi)$, while $\omega_{\text L}$ comes from LBD gravity. To illustrate the wormhole geometry, we will adopt the following setups for these parameters.

\begin{itemize}
  \item[(1)] Without any loss of generality, let $r_0=1$ for the throat radius, and $\phi_0=1$.

  \item[(2)] According to Eq.(\ref{Homogeneous scalar field ansatz}), asymptotic flatness of the spacetime requires $A>0$ so that the scalar field
      monotonically falls off as $r\to\infty$; moreover, $\phi(r)$ is positive definite and meets the expectation $G_{\text{eff}}=\phi^{-1}>0$ for the effective gravitational coupling strength, so that the graviton of LBD gravity is non-ghost for the sake of quantum stability.

  \item[(3)] A repulsive potential hill $V(\phi)>0$ tends to open and maintain the wormhole tunnel, while a trapping potential well $V(\phi)<0$ would collapse the wormhole tunnel. Thus, in our numerical modelings, let $V_0=1>0$ so that $V(\phi)$ serves as a potential hill.

  \item[(4)] Furthermore, it follows from Eq.(\ref{Throat constraint on parameters}) that the Lovelock parameter summarily satisfies
\begin{equation}\label{Throat constraint on parameters specific}
\begin{split}
\omega_{\text L}<-\frac{2}{A}<0\quad
\text{for}\;\; r_0=\phi_0=V_0=1\;\;\text{and}\;\;A>0\,.
\end{split}
\end{equation}
This agrees with the indication of Eqs.(\ref{Wormhole GEC Null}) and (\ref{Wormhole Weak GEC}) that a noncanonical ($\omega_{\text L}<0$) scalar field could best help support the wormhole.
\end{itemize}

\noindent With the numerical setups in Eq.(\ref{Throat constraint on parameters specific}), only two parameters $\omega_{\text L}$ and $A$ remain flexible in determining the behaviors of $b(r)$ and $(b-b'r)/b^2$, where  $A$ tunes the spatially decaying rate of the scalar field; appropriate values of  $\omega_{\text L}$ and $A$ should validate $b(r)<r$, $b-b'r>0$, and $\omega_{\text L}<-\frac{2}{A}<0$. In Fig.~\ref{Fig1}, $b(r)$ is plotted at the domain $r\geq r_0=1$, and the wormhole solution Eq.(\ref{Exact solution br}) is confirmed to be Lorentzian.
In Fig.~\ref{Fig2}, we plot $b-b'r$ and equivalently verify the flaring-out condition $(b-b'r)/b^2>0$. In both figures, we fix $A=2.3$ and illustrate the dependence on $\omega_{\text L}$ (note that inside Figs.~\ref{Fig1} $\sim$ \ref{Fig4}, $\omega_{\text L}$ is temporarily written as $\omega$ for the sake of greater clarity).




\vspace{-4mm}
\hspace{-10mm}
\begin{tabular}{c}
\vspace{-4mm}
\begin{minipage}{\linewidth}
 \makebox[\linewidth]{ \includegraphics[keepaspectratio=true,scale=0.44]{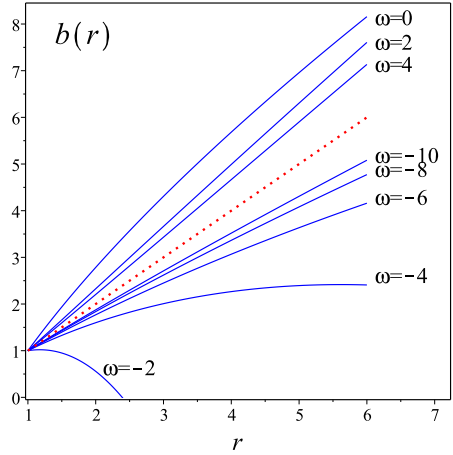}}
\vspace{-6cm}
\captionof{figure}{With $r_0=\phi_0=V_0=1$, $A=2.3$ and in the domain $r\geq r_0=1$, $b(r)$ is plotted as the solid curves for various $\omega_{\text L}$, along with the dotted diagonal for the auxiliary function $b(r)\equiv r$. For $\omega_{\text L}=\{-2,-4,-6,-8,-10\cdots\}<-2/A=-2/2.3$ in light of the numerical setups in Eq.(\ref{Throat constraint on parameters specific}), $b(r)$ always falls below the auxiliary diagonal $b(r)\equiv r$. Thus, $1-b(r)/r$ is positive definite and the wormhole solution Eq.(\ref{Exact solution br}) is Lorentzian. Moreover, the curve $b(r)$ approaches the dotted line when $\omega_{\text L}$ goes to $-\infty$, i.e. $\displaystyle \lim_{\omega_{\text L}\to-\infty}b(r)/r=1$. }\label{Fig1}
\end{minipage}
\\
\vspace{-4mm}
\begin{minipage}{\linewidth}
\makebox[\linewidth]{\includegraphics[keepaspectratio=true,scale=0.44]{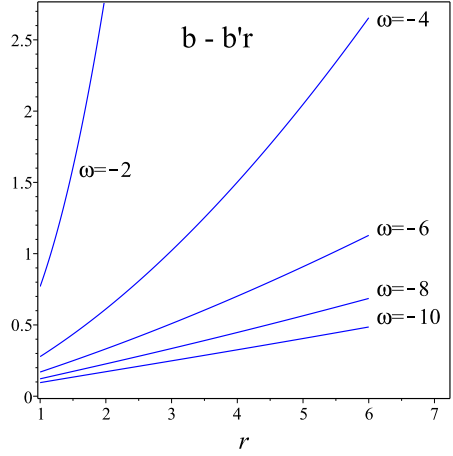}}
\vspace{-6cm}
\captionof{figure}{With $r_0=\phi_0=V_0=1$, $A=2.3$ and in the domain $r\geq r_0=1$,  $b-b'r$ is plotted for $\omega_{\text L}=\{-2,-4,-6,-8,-10\cdots\}<-2/A=-2/2.3$ and manifests itself to be positive definite. This equivalently confirms the outward-flaring condition $(b-b'r)/b^2>0$ of the embedding geometry. Moreover, the curve $b-b'r$ tends to coincide with the horizontal $r-$axis when $\omega_{\text L}$ approaches $-\infty$, i.e. $\displaystyle \lim_{\omega_{\text L}\to-\infty}b-b'r=0$, which is consistent with the tendency $\displaystyle \lim_{\omega_{\text L}\to-\infty}b(r)/r=1$ in Fig.~\ref{Fig1}.
}\label{Fig2}
\end{minipage}
\end{tabular}
\vspace{13.9mm}

\begin{figure}[h]
\begin{subfigure}{0.4\textwidth}
  \includegraphics[keepaspectratio=true,scale=0.36]{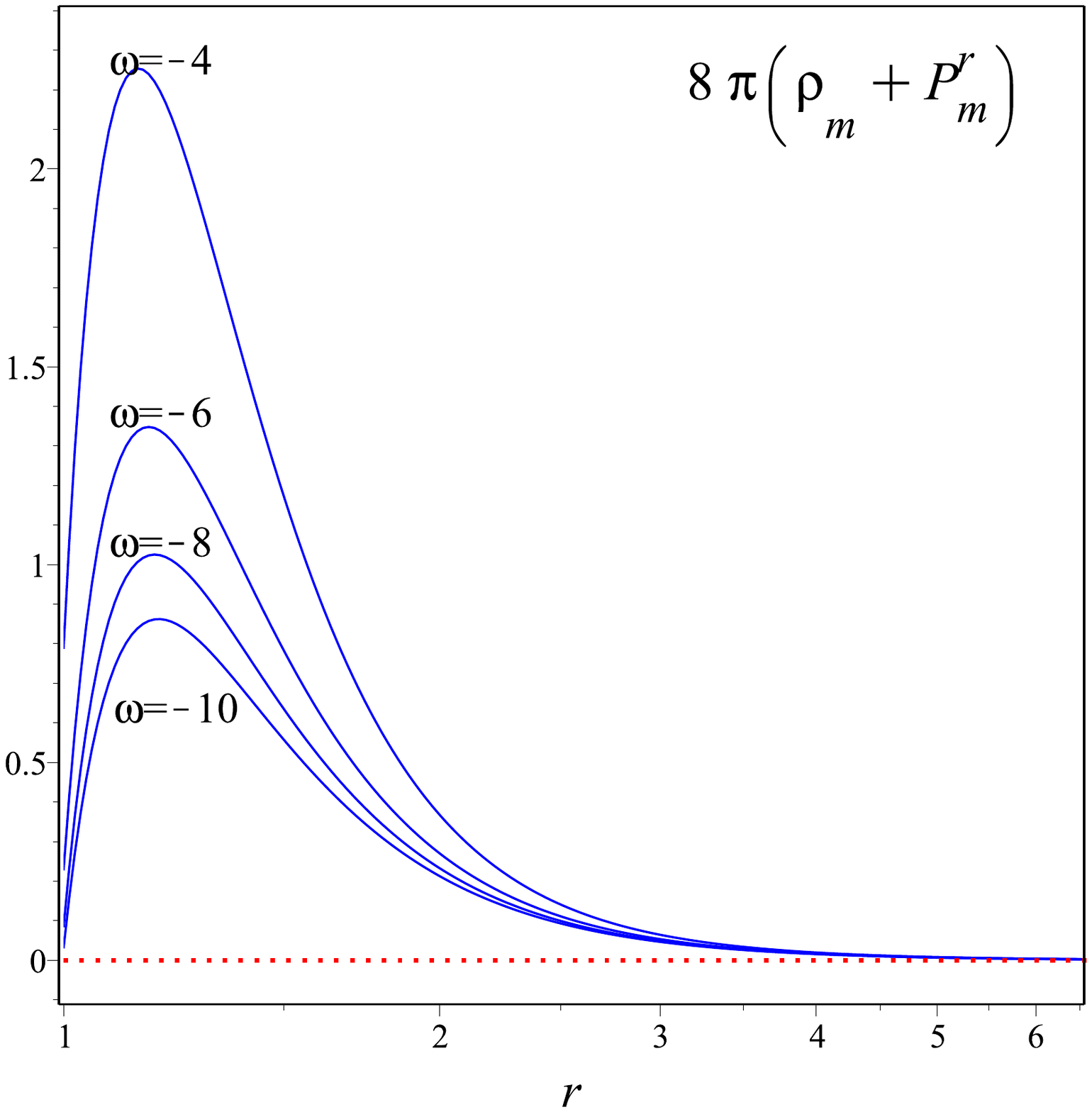}
\end{subfigure}
\hspace{5.6mm}
\begin{subfigure}{0.4\textwidth}
  \includegraphics[keepaspectratio=true,scale=0.36]{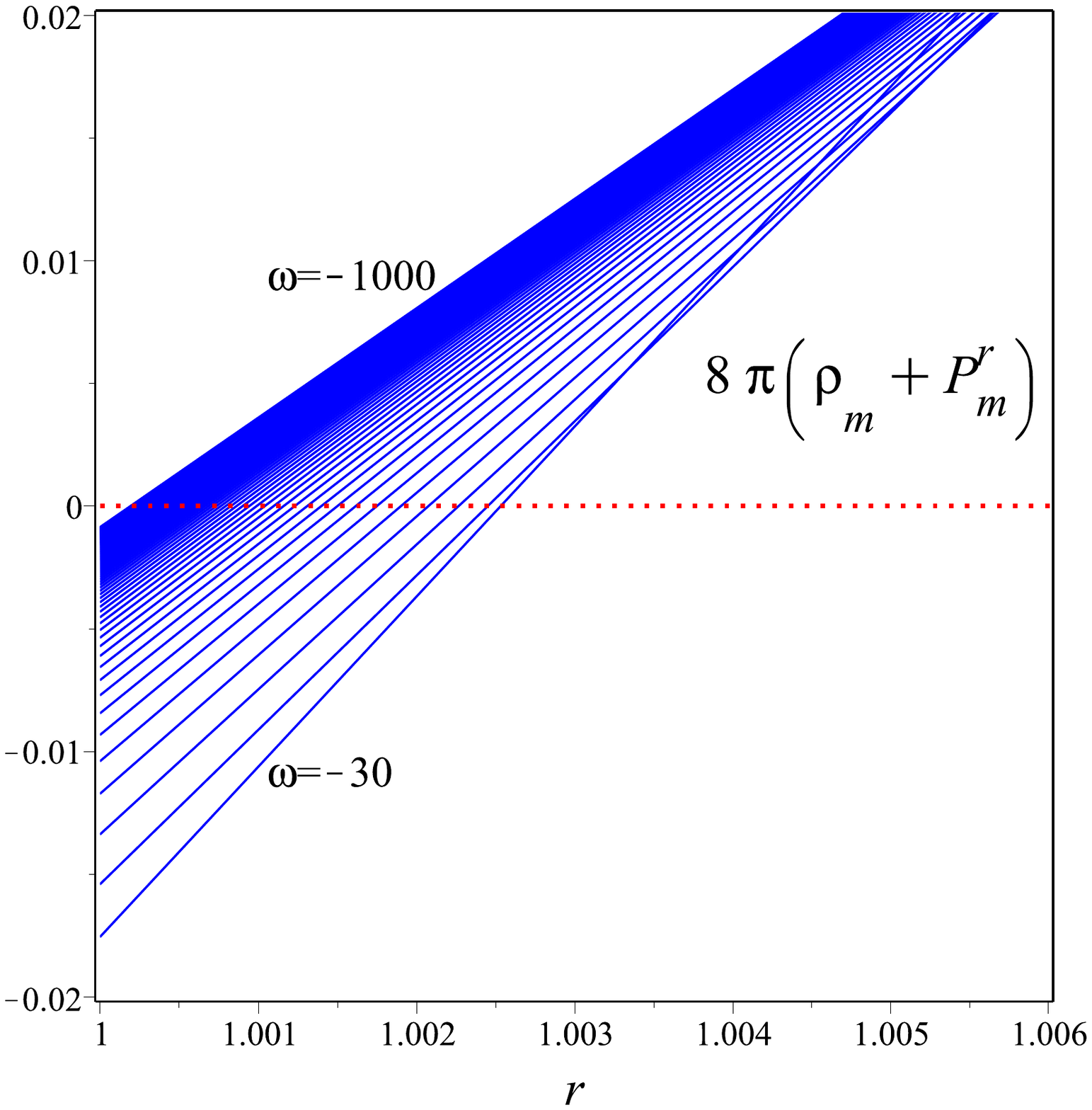}
\end{subfigure}
  \vspace{-3.2cm}
\caption{With $r_0=\phi_0=V_0=1$, $A=2.3$, $\hat{b}=-1$ and in the domain $r\geq r_0=1$, $8\pi(\rho_m +P_m^r)$ is plotted as the solid curves, while the dotted horizontal depicts the zero reference level. The first subfigure shows that for $\omega_{\text L}=\{-4,-6,-8,-10\}$, $8\pi(\rho_m +P_m^r)$ is positive definite with the expected asymptote $\displaystyle \lim_{r\to\infty}8\pi(\rho_m +P_m^r)=0^+$, so the standard NEC is respected. However, as $\omega_{\text L}$ further decreases, $8\pi(\rho_m +P_m^r)$ gradually falls below the dotted horizontal near the throat $r\gtrapprox  r_0=1$, which has been illustrated for $\omega_{\text L}=\{-30,-40,-50,\cdots,-1000\}$ in the second subfigure by magnifying the region $r_0=1\leq r\leq 1.0006$. Thus, large negative values of $\omega_{\text L}$ (numerical analysis gives $\omega_{\text L}\lesssim -12.9$) are unfavored in light of $8\pi(\rho_m +P_m^r)>0$.
}
\label{Fig3}
\end{figure}

\vspace{-7mm}
\begin{figure}
\begin{subfigure}{0.4\textwidth}
  \includegraphics[keepaspectratio=true,scale=0.36]{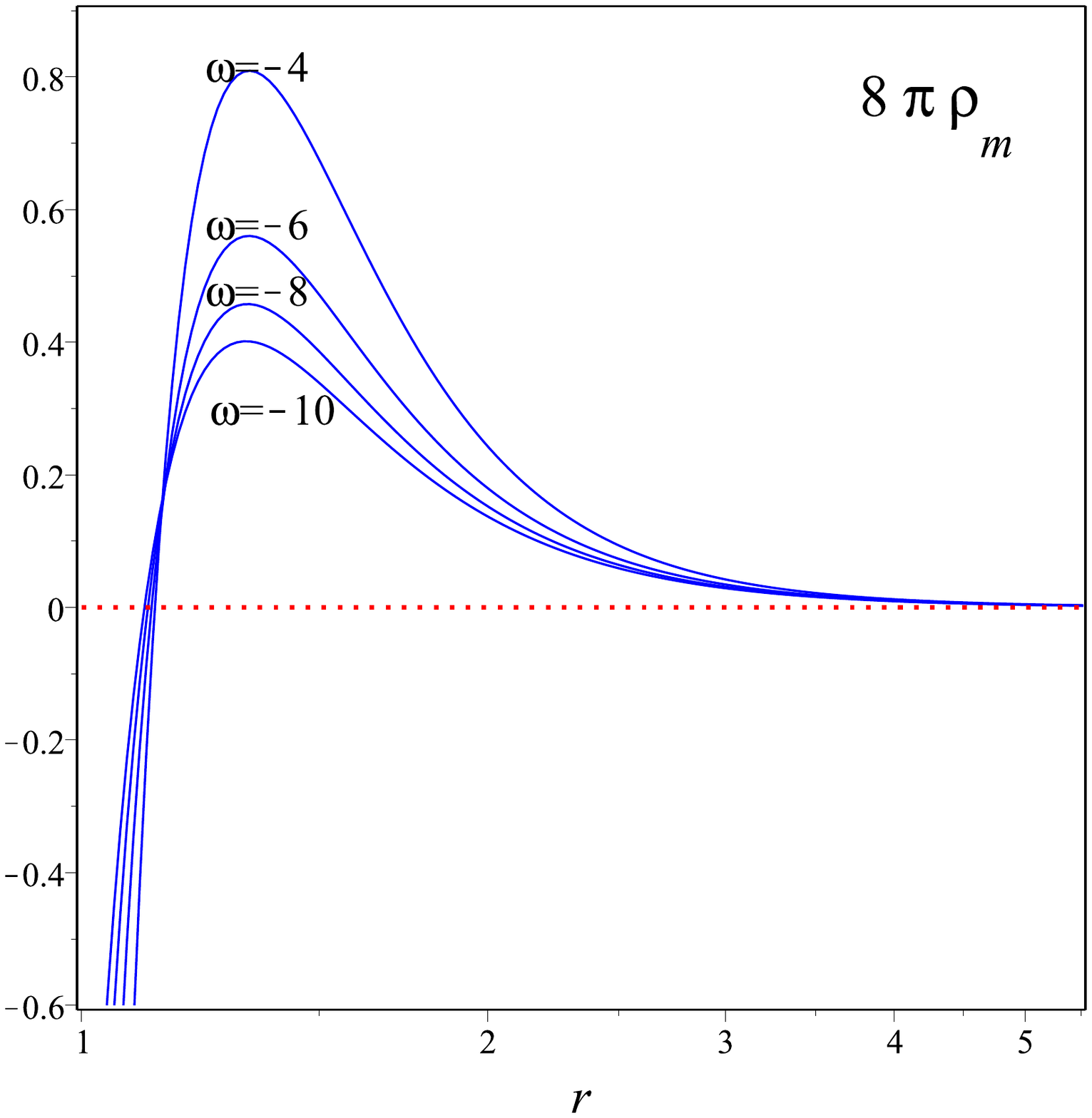}
\end{subfigure}
\hspace{5.6mm}
\begin{subfigure}{0.4\textwidth}
  \includegraphics[keepaspectratio=true,scale=0.36]{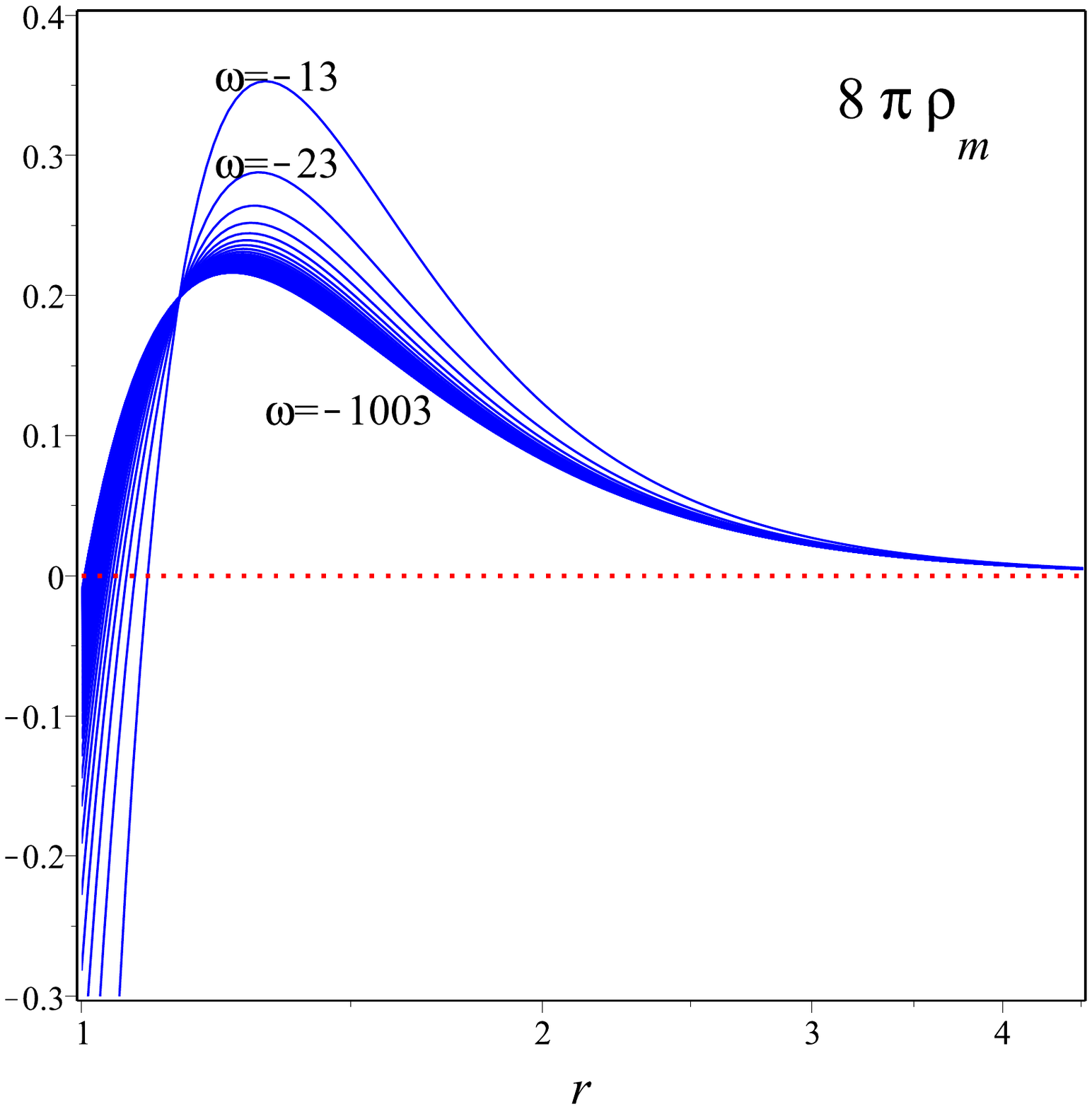}
\end{subfigure}
  \vspace{-3cm}
\caption{With $r_0=\phi_0=V_0=1$, $A=2.3$, $\hat{b}=-1$ and in the domain $r\geq r_0=1$, $8\pi\rho_m$ is plotted as the solid curves for $\omega_{\text L}=\{-4,-6,-8,-10\}$. Although $\rho_m>0$ in the distance, one always observes $\rho_m<0$ near the throat $r_0=1$, and thus the violation of the standard WEC, SEC and DEC. Moreover, as shown in the second subfigure for $\omega_{\text L}=\{-13,-23,-33,\cdots,-1003\}$, the intersection point between $8\pi\rho_m$ and the dotted zero reference level moves leftwards when $\omega_{\text L}$ decreases, so the violation of $\rho_m\geq 0$ gradually reduces. }\label{Fig4}
\end{figure}

\begin{figure}  \vspace{-16mm}
\begin{subfigure}{0.4\textwidth}
  \includegraphics[keepaspectratio=true,scale=0.36]{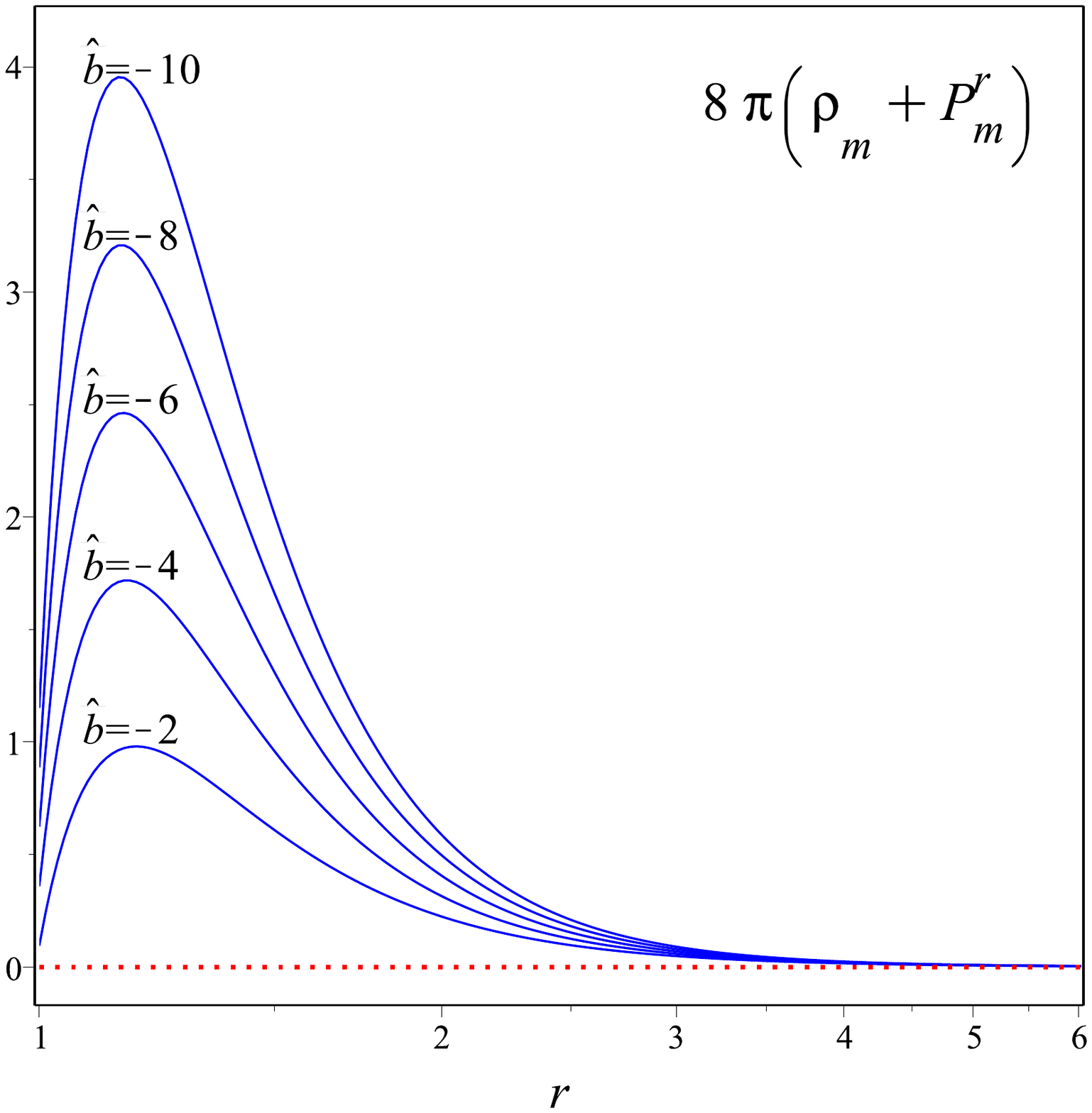}
\end{subfigure}
\hspace{5.6mm}
\begin{subfigure}{0.4\textwidth}
  \includegraphics[keepaspectratio=true,scale=0.36]{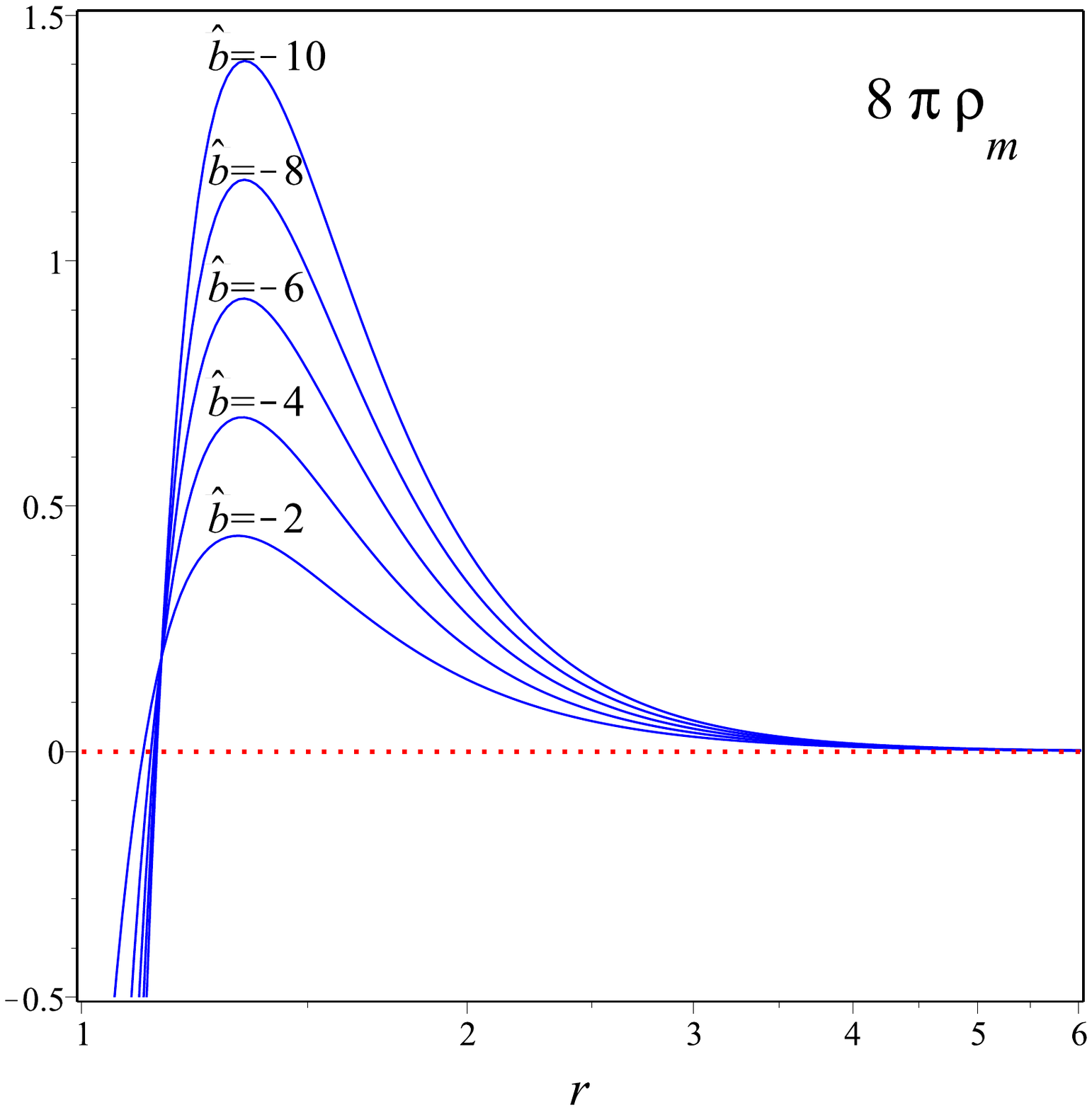}
\end{subfigure}
  \vspace{-3cm}
\caption{With $r_0=\phi_0=V_0=1$, $\omega_{\text L}=-6$, $A=2.3$ and in the domain $r\geq r_0=1$, we plot $8\pi(\rho_m +P_m^r)$ and $8\pi\rho_m$ for different Gauss-Bonnet topology-gravity coupling strength, as is given the decreasing series $\hat b=\{-2,-4,-6,-8,-10\}$. The standard NEC always holds with $\rho_m +P_m^r\geq 0$. Moreover, the intersection point between $8\pi\rho_m$ and the dotted reference level moves leftwards when $\hat b<0$ increases from $\hat b=-10$ to $-2$,, so the violation of $\rho_m\geq 0$ gradually reduces.}
\label{Fig5}
\end{figure}

\begin{figure}\vspace{-6mm}
\begin{subfigure}{0.4\textwidth}
  \includegraphics[keepaspectratio=true,scale=0.36]{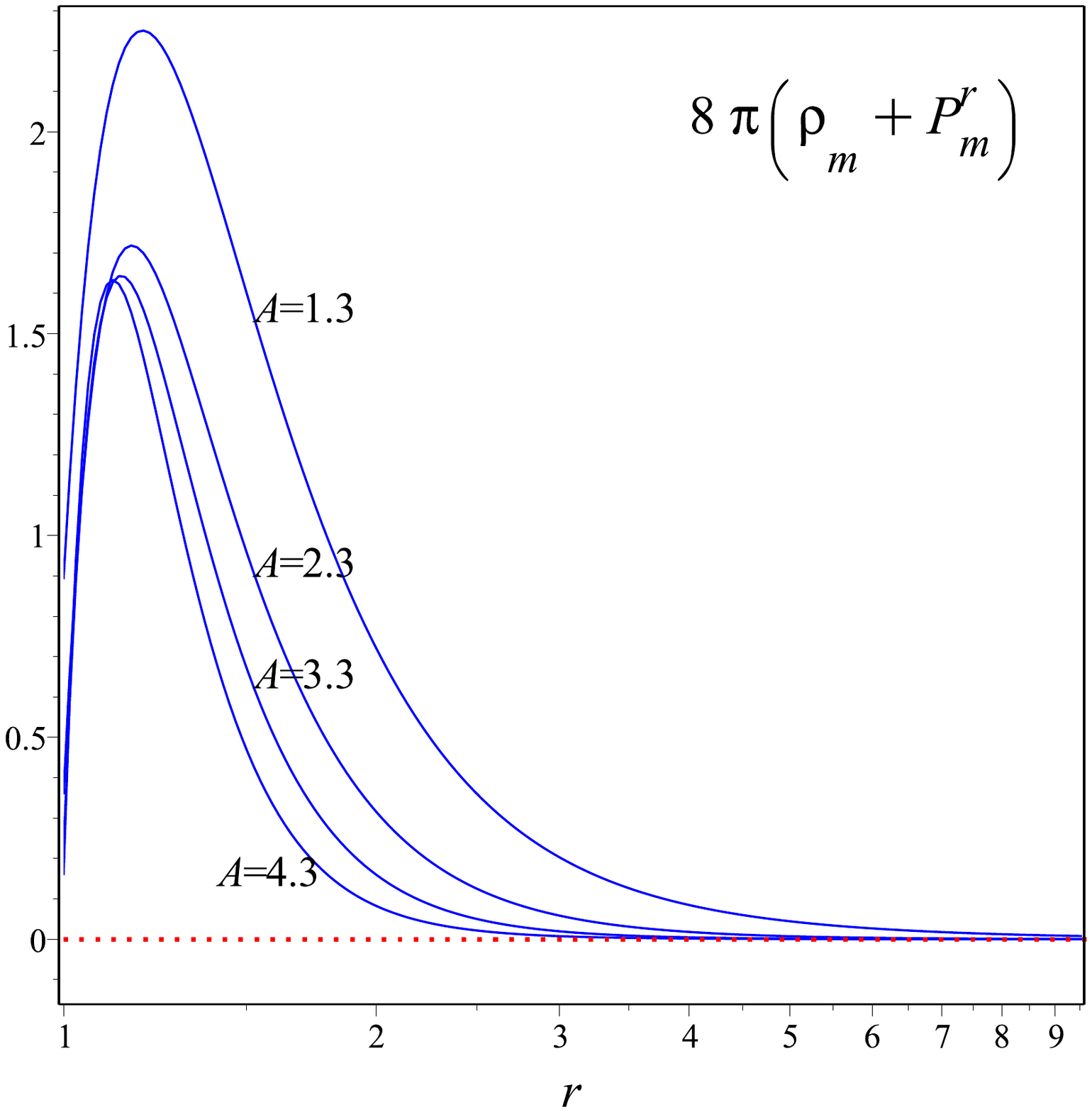}
\end{subfigure}
\hspace{5.6mm}
\begin{subfigure}{0.4\textwidth}
  \includegraphics[keepaspectratio=true,scale=0.36]{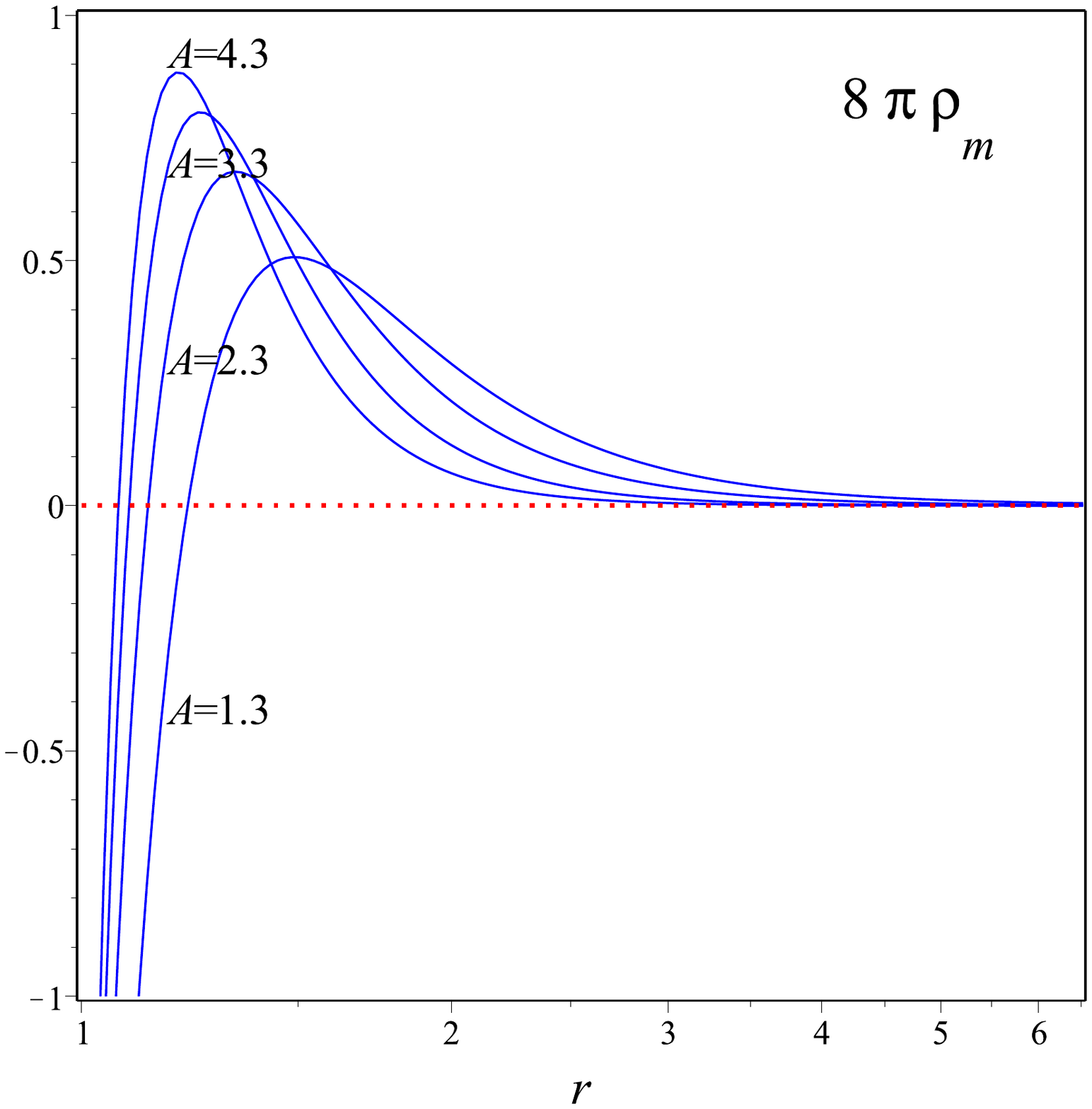}
\end{subfigure}
  \vspace{-3.2cm}
\caption{With $r_0=\phi_0=V_0=1$, $\omega_{\text L}=-6$, $\hat b=-4$ and in the domain $r\geq r_0=1$, we plot $8\pi(\rho_m +P_m^r)$ and $8\pi\rho_m$ for different decaying rate of the scalar field, as is given by the increasing series $A=\{1.3,2.3,3.3,4.3\}$.  The standard NEC always holds with $\rho_m +P_m^r\geq 0$. Moreover, the intersection point between $8\pi\rho_m$ and the dotted reference level moves leftwards when $A>0$ increases from $A=1.3$ to $4.3$, so the violation of $\rho_m\geq 0$ gradually reduces.}
\label{Fig6}
\end{figure}

With the Einstein tensor $G^\mu_{\;\,\nu}$ given by Eq.(\ref{curvature invariants and Einstein tensor}) and $\phi^{-1}>0$, adding up the componential field equations $G^t_{\;\,t}=-8\pi\phi^{-1}\rho_{\text{eff}}$ and $G^r_{\;\,r}=8\pi\phi^{-1}P^r_{\text{eff}}$, one could obtain $b-b'r=-r^3\cdot 8\pi\phi^{-1} (\rho_{\text{eff}}+P^r_{\text{eff}} )$. Thus, for the numerical setups summarized by Eq.(\ref{Throat constraint on parameters specific}), Fig.~\ref{Fig2} not only verifies the positive definiteness of $b-b'r$, but also implies the violation of the GNEC $\phi^{-1}(\rho_{\text{eff}}+P^r_{\text{eff}})<0$ -- and consequently the GWEC, GSEC and GDEC in LBD gravity.
On the other hand, can the standard energy conditions in Eq.(\ref{GR Null and Stong ECs}) still hold along the radial direction for the matter threading the wormhole? The energy density $\rho_m$ and the radial pressure $P^r_m$ vary for different types of physical matter, and $\rho_m+P^r_m$ relies on the the equation of state $P^r_m=P^r_m(\rho_m)$. Thus, we choose to calculate  $\rho_m+P^r_m$ from an indirect approach. Considering that $\frac{\phi}{8\pi}G^t_{\;\,t}=- (\rho_m+\rho_\phi+a\rho_{\text{CP}}+\hat{b}\rho_{\text{GB}})$ and $\frac{\phi}{8\pi}G^r_{\;\,r}=P_m^r+P_\phi^r+aP_{\text{CP}}^r
+\hat{b}P_{\text{GB}}^r$, $\rho_m$ and $P^r_m$ can be recovered by
\begin{equation}
\begin{split}
8\pi \rho_m= b'r\phi -  8\pi\rho_\phi  -  8\pi \hat{b}\rho_{\text{GB}} \quad,\quad
8\pi P^r_m=-b\phi  -  8\pi P^r_\phi  -  8\pi \hat{b}P^r_{\text{GB}} \,,
\end{split}
\end{equation}
where, according to Eqs.(\ref{Generic rho P 1})-(\ref{Generic rho P 4}) with $\Phi(r)=0$, we have
\begin{align}
8\pi\rho_\phi= &\,\left(1-\frac{b}{r}\right) \left(\phi''+\frac{2\phi'}{r}+\frac{\omega_{\text L}}{2} \frac{\phi'^2}{\phi} \right)
+\frac{\phi'}{2r^2}\left( b- b'r \right) +V \\
&8\pi P^r_{\phi}= \left( 1 -\frac{b }{ r}\right)\left(\phi''+\omega_{\text L}\frac{ \phi'^2 }{\phi } \right) -8\pi\rho_\phi\\
8\pi\rho_{\text{GB}} &= \frac{2}{r^5} \Big[\phi' \left(b-b'r\right)\left(3b -2r \right) + 2br\phi''(r-b)  \Big]\\
&\mbox{and}\quad 8\pi P^r_{\text{GB}}= \frac{2\phi'}{r^4} (b-b'r)(b'-2)\,.
\end{align}

In Fig.~\ref{Fig3}, $8\pi(\rho_m +P_m^r)$ is plotted as the solid curves, where we let $\hat b=-1<0$ for the Gauss-Bonnet matter-topology coupling strength so that the Gauss-Bonnet part of LBD gravity could yield antigravitational effect to help maintain the wormhole tunnel.
$\rho_m +P_m^r$ is positive definite for $\omega_{\text L}=\{-4,-6,-8,-10\}<-2/A$ and thus the standard NEC $\rho_m +P^r_m\geq 0$ is respected by the physical matter, despite the violation of the GNEC due to $\phi^{-1}(\rho_{\text{eff}}+P^r_{\text{eff}})<0$; in fact, this has realized the null-energy supporting condition of Eq.(\ref{Wormhole GEC Null}) in an anisotropic perfect fluid form. However, large negative values of $\omega_{\text L}$ is unfavored: careful numerical analysis finds that the standard NEC becomes slightly violated, i.e. $\rho_m +P^r_m<0$  for $\omega_{\text L}\lesssim -12.9$ in the very close vicinity of the wormhole throat.


Validity of the standard weak, strong and dominant energy conditions requires us to check the positivity of the physical matter density $\rho_m$.
Plotting $8\pi\rho_m$ for $\omega_{\text L}=\{-4,-6,-8,-10\cdots\}$
in Fig.~\ref{Fig4}, we find $\rho_m<0$ near the wormhole throat, and
the violation of $\rho_m\geq 0$ can be reduced with the decrement of $\omega_{\text L}$ in the domain $\omega_{\text L}<-2/A$; actually, this has negated the weak-energy supporting condition of Eq.(\ref{Wormhole Weak GEC}) in the anisotropic perfect fluid form, despite the validity of the null-energy Eq.(\ref{Wormhole GEC Null}). As expected, when one goes way from the wormhole throat, the normal behaviors $\rho_m\geq 0$ and $\displaystyle \lim_{r\to\infty} \rho_m =0^+$ are recovered, and thus the standard WEC becomes valid as $\rho_m +P^r_m\geq 0$ for $r>r_0=1$ in light of Fig.~\ref{Fig3}.


Having seen from Fig.~\ref{Fig4} that the decrement of the noncanonical $\omega_{\text L}$ could reduce the violation of $\rho_m\geq 0$, we cannot help but ask are there any other factors that could help protect the standard WEC? The answer is yes. In Figs.~\ref{Fig5} and \ref{Fig6}, we respectively fix $\{\omega_{\text L}=-6\,, A=2.3\}$ and $\{\omega_{\text L}=-6\,, \hat b=-4\}$ to plot $\{8\pi(\rho_m +P_m^r)\,,8\pi\rho_m\}$. It turns out that when the standard NEC is obeyed, i.e. $\rho_m +P_m^r\geq0$, the increment of $\hat b$ (Gauss-Bonnet topology-gravity coupling strength) in the repulsive domain $\hat b<0$ and the increment of $A$ (decaying-rate index of the scalar field) in the domain $A>0$ could both help minimize the violation of $\rho_m\geq 0$ near the wormhole throat.

\section{Implication: Wormholes in Brans-Dicke gravity}\label{Sec Wormholes in Brans-Dicke gravity}

In the limits $a\to 0$ and $\hat b \to 0$ for the parity-gravity and the topology-gravity coupling coefficients in $\mathcal{S}_{\text{LBD}}$, and in the absence of the potential $V(\phi)$, LBD gravity reduces to become Brans-Dicke gravity with the standard action \cite{Brans Dicke gravity}
\begin{equation}\label{Action BD gravity}
\begin{split}
\mathcal{S}_{\text{BD}}=\frac{1}{16\pi}\int d^4x\sqrt{-g}\,\left( \phi R  -\frac{\hat\omega }{\phi}\nabla_\alpha \phi
\nabla^\alpha\phi 
\right)+ \mathcal{S}_m\,,
\end{split}
\end{equation}
where $\hat\omega$ refers to the Brans-Dicke parameter (in distinction with $\omega_{\text L}$ for the Lovelock parameter). The gravitational field equation $\delta\mathcal{S}_{\text{BD}}/\delta g^{\mu\nu}=0$ and the kinematical wave equation $\delta\mathcal{S}_{\text{BD}}/\delta\phi=0$ are respectively
\begin{equation}\label{BD field equation}
\begin{split}
\phi\left(R_{\mu\nu} -\frac{1}{2}  R  g_{\mu\nu}\right)
+ \left(g_{\mu\nu}\Box -\nabla_\mu \nabla_\nu \right) \phi-\frac{\hat\omega }{\phi}\left(\nabla_\mu \phi \nabla_\nu \phi-\frac{1}{2}g_{\mu\nu}\nabla_\alpha \phi
\nabla^\alpha\phi\right)
=8\pi  T_{\mu\nu}^{\text{(m)}}\,,
\end{split}
\end{equation}
\begin{equation}\label{kinematic wave equation for BD}
\text{and}\quad
2\hat\omega \cdot \Box\phi  =
-\phi R+\frac{ \hat\omega }{\phi}
\nabla_{\alpha}\phi \nabla^{\alpha} \phi
\,.
\end{equation}
With the trace of the field equation $-\phi R+\frac{\hat\omega }{\phi}\nabla_\alpha \phi \nabla^\alpha\phi+3 \Box \phi
=8\pi  T^{\text{(m)}}$, Eq.(\ref{kinematic wave equation for BD}) leads to the dynamical wave equation
$
(2\hat\omega  +3)\Box\phi =  8\pi T^{\text{(m)}}
$; however, the kinematical equation (\ref{kinematic wave equation for BD}) is preferred so that temporarily we need not worry about $T^{\text{(m)}}$ for the physical matter.

\vspace{-3mm}
\begin{figure}[h]
\hspace{-3mm}
\begin{subfigure}{0.4\textwidth}
  \includegraphics[keepaspectratio=true,scale=0.37]{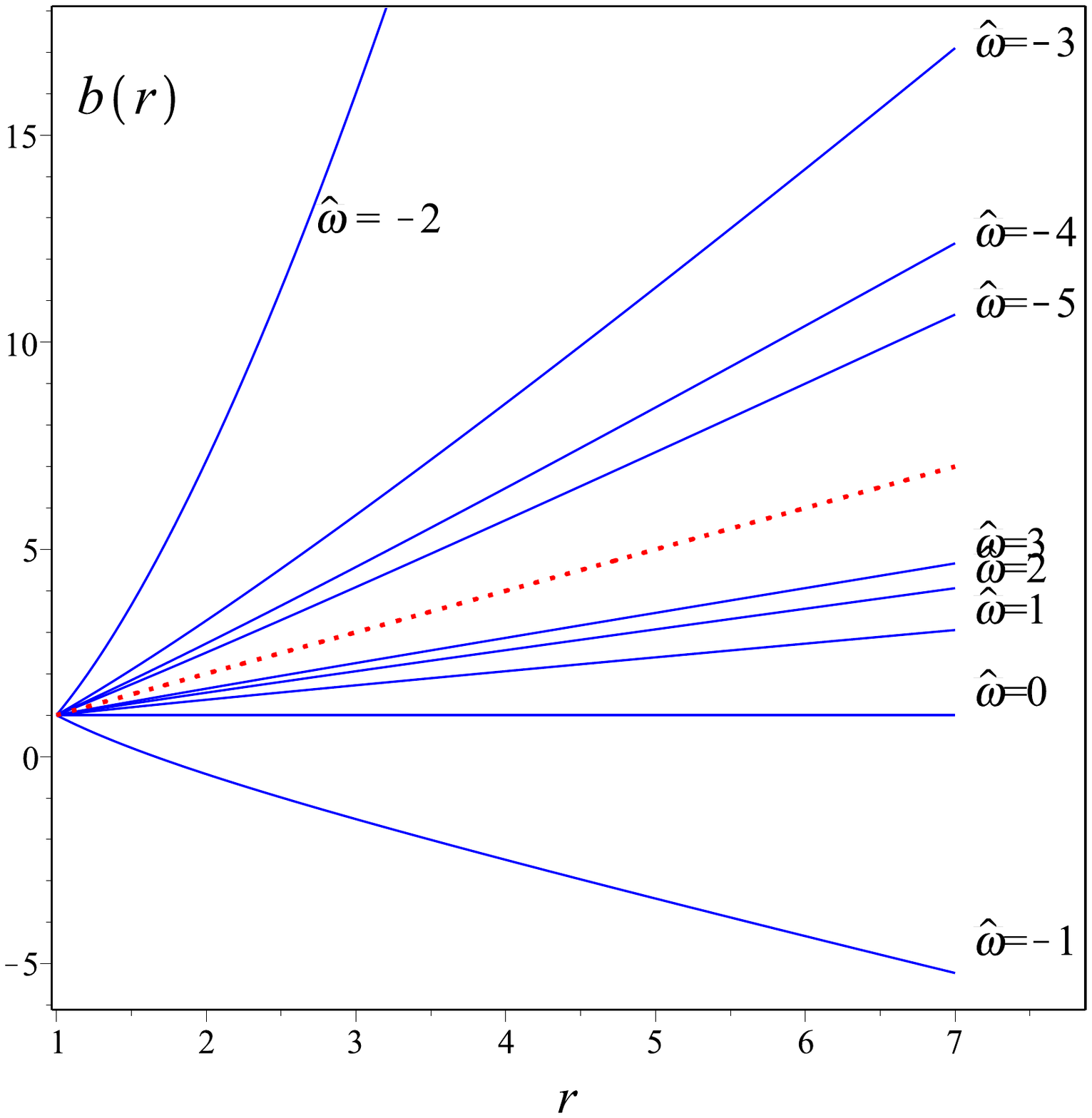}
\end{subfigure}
\hspace{8mm}
\begin{subfigure}{0.4\textwidth}
  \includegraphics[keepaspectratio=true,scale=0.37]{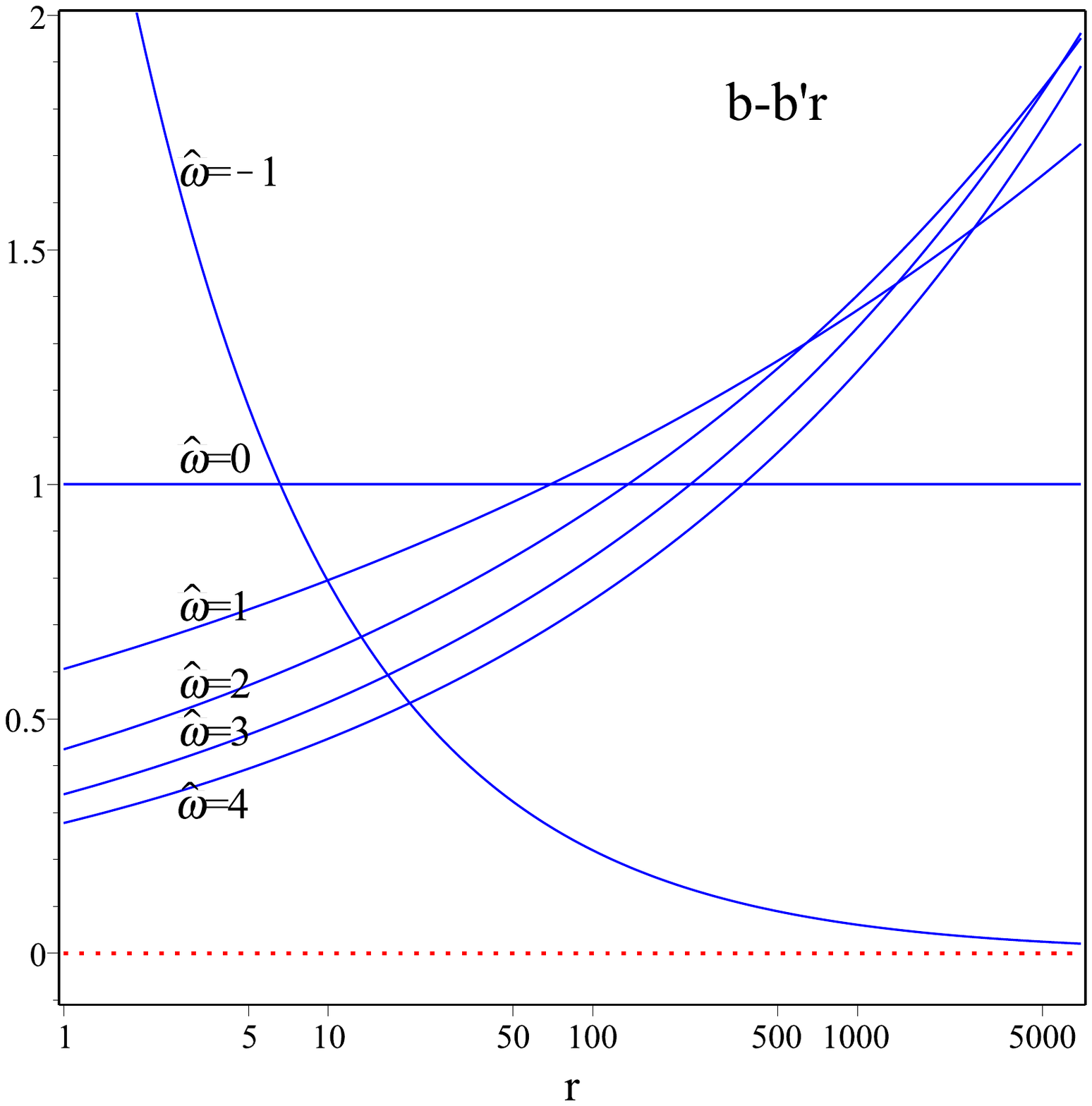}
\end{subfigure}
  \vspace{-3.3cm}
\caption{With $r_0=\phi_0=1$ and $A=1.3$, we plot $b(r)$ and $b-b'r$ as the solid curves. The first subfigure shows that for $\hat\omega=\{-1,0,1,2,3\cdots\}>-2/A=-2/1.3$, $b(r)$ always falls below the dotted diagonal of the auxiliary function $b(r)\equiv r$, and thus guarantees the Lorentzian signature as $1-b(r)/r>0$. Moreover, the second subfigure verifies $b-b'r>0$ for $\hat\omega=\{-1,0,1,2,3,4\cdots\}>-1/1.3$, so the flaring-out condition $(b-b'r)/b^2>0$ of the embedding geometry is satisfied.}
\label{Fig7}
\end{figure}

For Morris-Thorne wormholes in Brans-Dicke gravity, consider a zero-tidal-force solution $\Phi(r)=0$, and inherit the ansatz $\phi(r)=\phi_0\left(\frac{r_0}{r} \right)^ A$ ($\phi_0>0$, $ A=\text{constant}$) of Eq.(\ref{Homogeneous scalar field ansatz}) for the scalar field. Directly solving Eq.(\ref{kinematic wave equation for BD}) for $b(r)$ with the boundary condition $b(r_0)=r_0$, or just substituting $\{V_0\equiv0, \omega_{\text L}\mapsto\hat\omega\}$ into Eqs.(\ref{Exact solution br}) and (\ref{Exact solution b-br'}), we obtain
\begin{equation}\label{BD wormhole}
b(r)=r+
\frac{2r - 2r_0 \left(\frac{r_0}{r}\right)^{\frac{\hat\omega A(1- A)}{\hat\omega A +2}}  }
{\hat\omega A^2 -2\hat\omega A -2} \;\;\quad\mbox{and}\;\;\quad
b-b'r=\frac{2 r_0}
{\hat\omega A+2 }
\left(\frac{r_0} {r }\right)^{\frac{\hat\omega  A(1- A)}{\hat\omega A +2}} \,.
\end{equation}
In light of the flaring-out condition at the wormhole throat, the parameters $\{A,\hat\omega\}$ have to meet the requirement
\begin{equation}\label{BD Throat constraint on parameters}
b'(r_0)
=1-\frac{2}{\hat\omega A +2}<1
\quad\Rightarrow\quad \hat\omega A>-2
\,.
\end{equation}
Note that this condition does not conflict with the $\omega_{\text L} A<-2$ in Eq.(\ref{Throat constraint on parameters specific}): Eq.(\ref{BD Throat constraint on parameters}) comes from Eq.(\ref{Throat constraint on parameters}) with  $\{V_0\equiv0, \omega_{\text L}\mapsto\hat\omega\}$, while Eq.(\ref{Throat constraint on parameters specific}) specifies Eq.(\ref{Throat constraint on parameters}) by $r_0=\phi_0=V_0=1$; the choices $V_0\equiv0$ and $V_0=1$ (and also $V_0=-1$, if one would like to check it), i.e. the potential being vanishing, repulsive or attractive, lead Eq.(\ref{Throat constraint on parameters}) to totally different situations.

\vspace{-3mm}
\hspace{-9mm}
\begin{minipage}{\linewidth}
\makebox[\linewidth]{
  \includegraphics[keepaspectratio=true,scale=0.42]{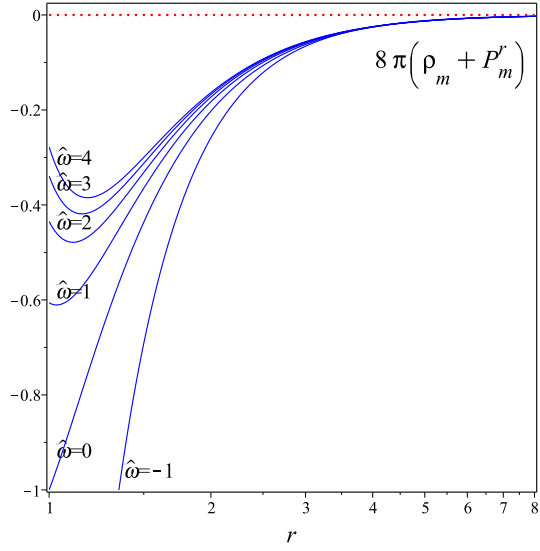}}
  \vspace{-4cm}
\captionof{figure}{With $r_0=\phi_0=1$ and $A=1.3$, we plot $8\pi(\rho_m+P^r_m)$ for $\hat\omega=\{-1,0,1,2,3,4\}$ as the solid curves, which always fall below the dotted horizontal for the zero reference level. Thus, the standard NEC is always violated. As $\hat\omega$ grows from -1 to 4, the curve of  $8\pi(\rho_m+P^r_m)$ gradually moves upward, so in a sense the violation of $8\pi(\rho_m+P^r_m)\geq0$ can be reduced for greater values of $\hat\omega$ in the domain $\hat\omega>-2/A$.}\label{Fig8}
\end{minipage}\\

\vspace{-3mm}
\begin{figure}[h]
\hspace{-4mm}
\begin{subfigure}{0.4\textwidth}
  \includegraphics[keepaspectratio=true,scale=0.4]{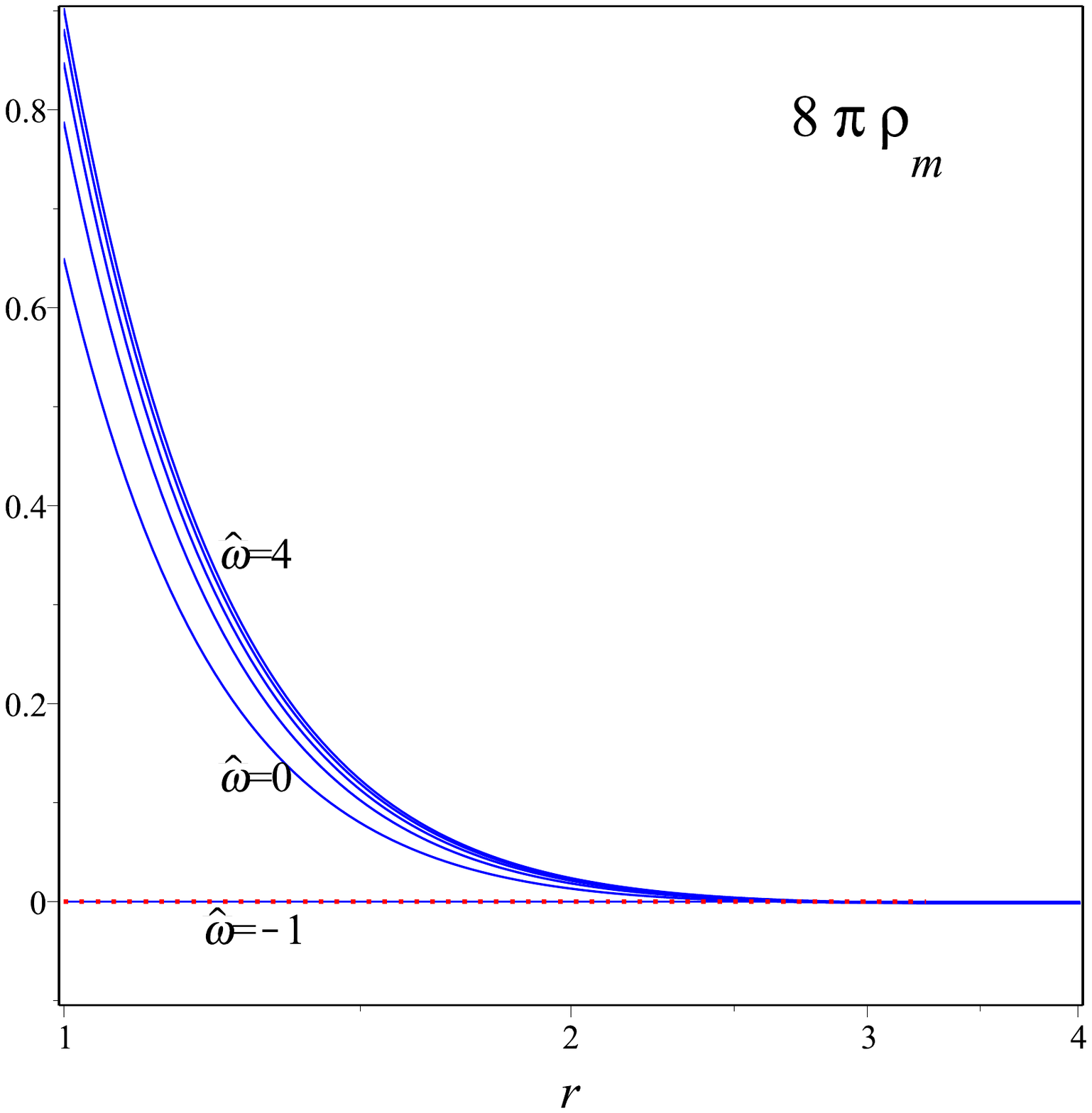}
\end{subfigure}
\hspace{9mm}
\begin{subfigure}{0.4\textwidth}
  \includegraphics[keepaspectratio=true,scale=0.4]{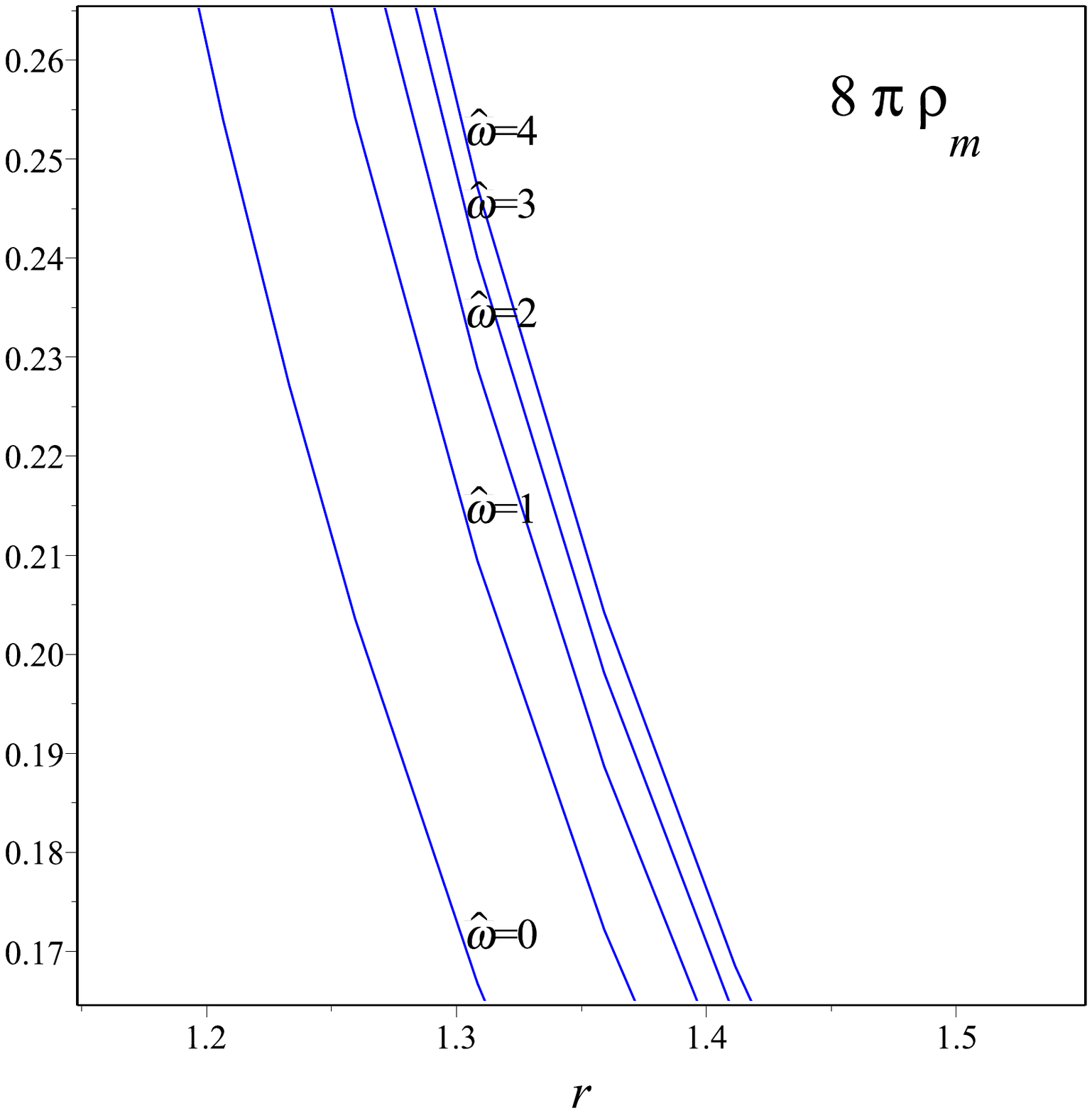}
\end{subfigure}
  \vspace{-3.4cm}
\caption{With $r_0=\phi_0=1$ and $A=1.3$, we plot $8\pi \rho_m$ for $\hat\omega=\{-1,0,1,2,3,4\}$ as the solid curves. These curves stay above the zero reference level for $\hat\omega=\{0,1,2,3,4\}$ and coincide with it for $\hat\omega=-1$; since the curves for $\hat\omega=\{0,1,2,3,4\}$ are stickily close to each other in the first subfigure, we magnify them at $1.2<r<1.45$ for greater clarity in the second subfigure, which shows that from bottom to top or from left to right, the curves correspond to $\hat\omega=0,...4$ in sequence. Although the energy density is nonnegative for $\hat\omega\geq-1$, the standard WEC, SEC and DEC still fail as $8\rho_m+P^r_m<0$ for all  $\hat\omega>-2/A$ by Fig.~\ref{Fig8}. }
\label{Fig9}
\end{figure}



Among the three parameters in $b(r)$, $\{\hat\omega\,, A\}$ jointly determine the wormhole structure, while $r_0$ acts as an auxiliary parameter; for the same reasons indicated in the proceeding section, we inherit the numerical setups $\{r_0=\phi_0=1,A>0\}$ to illustrate the Brans-Dicke wormhole Eq.(\ref{BD wormhole}), which implies $\hat\omega>-2/A$ from Eq.(\ref{BD Throat constraint on parameters}). To start with, the Lorentzian-signature condition $r>b(r)$ and the outward-flaring constraints $b-b'r>0$ are confirmed in Fig.~\ref{Fig7}.

Next, let's check the states of the standard energy conditions in Eq.(\ref{GR Null and Stong ECs}). For the matter threading the wormhole, the energy density and radial pressure can be indirectly reconstructed from the field equations
$8\pi \rho_m= b'r\phi -  8\pi\rho_\phi$ and $8\pi P^r_m=-b\phi-8\pi P^r_\phi$, where
\begin{equation}
8\pi\rho_\phi= \left(1-\frac{b}{r}\right) \left(\phi''+\frac{2\phi'}{r}+\frac{\hat\omega}{2} \frac{\phi'^2}{\phi} \right)
+\frac{\phi'}{2r^2}\left( b- b'r \right) \,,
\end{equation}
\begin{equation}
8\pi P^r_{\phi}=
\left( 1 -\frac{b }{ r}\right)\left(\phi''+\hat\omega\frac{ \phi'^2 }{\phi } \right) -8\pi\rho_\phi\,.
\end{equation}
With $\phi_0=1=r_0$ in $\phi(r)=\phi_0\left(\frac{r_0}{r} \right)^A$ and the zero-tidal-force solution Eq.(\ref{BD wormhole}), it follows that
\begin{equation}\label{BD rhom+Pm}
\begin{split}
8\pi (\rho_m+P^r_m)=\frac{2\hat\omega A\big(\hat\omega A^2+ A^2
+4 A-2\big)
+4\big( A^2+ A-1\big)}{(\hat\omega A+2)(\hat\omega A^2 -2\hat\omega A -2)}\frac{r^{\frac{\hat\omega  A( A-1)}{\hat\omega A +2}}}{r^{ A+3}}
-\frac{2 A(\hat\omega A+ A+1)}{(\hat\omega A^2 -2\hat\omega A -2)}\frac{r}{r^{ A+3}}\,,
\end{split}
\end{equation}
\begin{equation}\label{BD rhom}
\begin{split}
8\pi\rho_m=\frac{1}{r}+
\frac{r(\hat\omega+6)-\left[\hat\omega+3
+\frac{3\hat\omega A( A-1)}{\hat\omega A+2}\right]r^{\frac{\hat\omega A( A-1)}{\hat\omega A +2}}}
{r^2(\hat\omega A^2-2\hat\omega A-2)}\,.
\end{split}
\end{equation}

\begin{center}
\begin{figure}
\hspace{10mm}
\begin{subfigure}{0.4\textwidth}\hspace{-16mm}
  \includegraphics[keepaspectratio=true,scale=0.415]{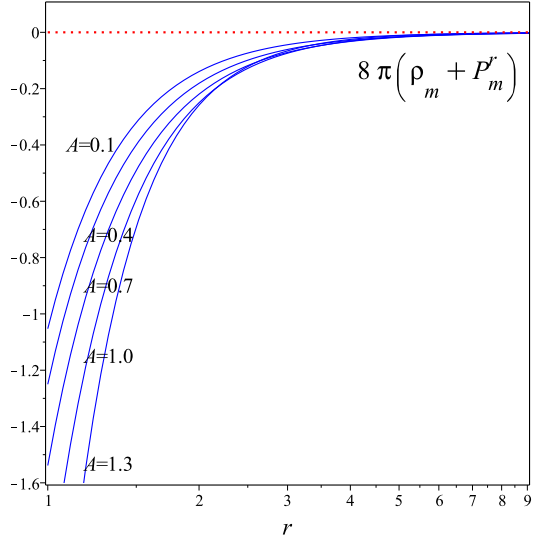}
  \vspace{-40mm}\hspace{30mm}\caption{From top to bottom, $A=$0.1, 0.4, $\cdots$, 1.3.}
\label{Fig10a}
\end{subfigure}\qquad\qquad
\begin{subfigure}{0.4\textwidth}\hspace{-16mm}
  \includegraphics[keepaspectratio=true,scale=0.415]{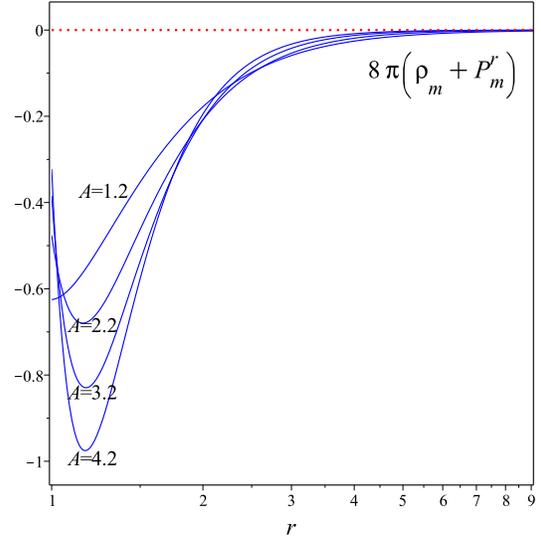}
  \vspace{-40mm}\caption{From top to bottom, $A=$1.2, 2.2, 3.2, 4.2.}
\label{Fig10b}
\end{subfigure}

\hspace{10mm}
\begin{subfigure}{0.4\textwidth}\hspace{-16mm}
  \includegraphics[keepaspectratio=true,scale=0.415]{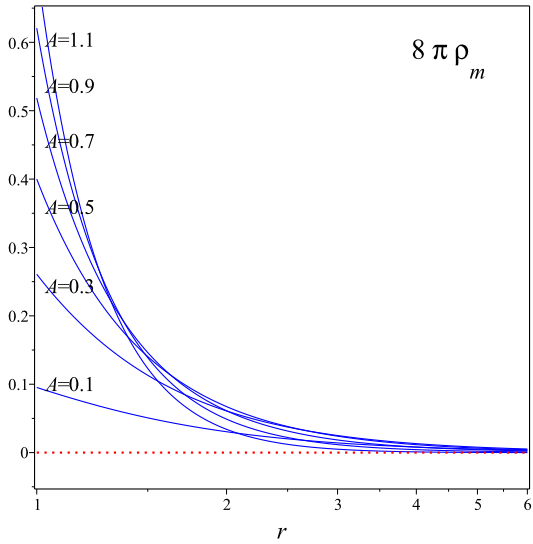}
  \vspace{-40mm}\caption{From top to bottom, $A=$1.1, 0.9, $\cdots$, 0.1.}
\label{Fig10c}
\end{subfigure}\qquad\qquad
\begin{subfigure}{0.4\textwidth}\hspace{-16mm}
  \includegraphics[keepaspectratio=true,scale=0.415]{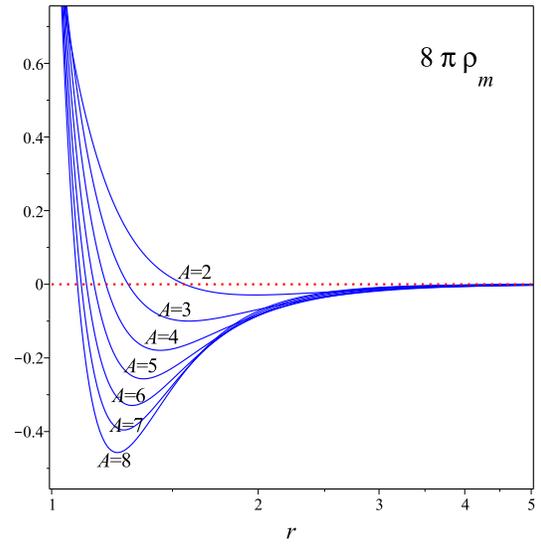}
  \vspace{-40mm}\caption{From top to bottom, $A=$2, 3, $\cdots$, 8.}
\label{Fig10d}
\end{subfigure}
  \vspace{5mm}
\caption{With $r_0=\phi_0=1$ and $\hat\omega=1$, we plot $8\pi (\rho_m+P^r_m)$ and $8\pi \rho_m$ for different values of $A$ which governs the spatially decaying rate of the scalar field. Figs.~\ref{Fig10a} and \ref{Fig10b}, with a similar appearance to Fig.~\ref{Fig8}, show $\rho_m+P^r_m<0$ for $A>0$ so that the standard NEC is violated; moreover, with Fig.~\ref{Fig10c} in an analogous pattern to Fig.~\ref{Fig9}, one finds $\rho_m>0$ for slow decaying rate $A=\{0.1,0.3,\cdots,1.1\}$. As the most interesting observation,  Fig.~\ref{Fig10d} shows that $\rho_m$ is no longer positive definite throughout $r\geq r_0$ for high decaying rate $A\gtrsim2$.}
\label{Fig10}
\end{figure}
\end{center}

\noindent Based on Eqs.(\ref{BD rhom+Pm}) and (\ref{BD rhom}), the behaviors of $8\pi (\rho_m+P^r_m)$ and $8\pi \rho_m$ are illustrated in Figs.~\ref{Fig8} and \ref{Fig9} with the numerical setups $\{r_0=\phi_0=1,A=1.3>0,\hat\omega>-2/A\}$. Unfortunately, despite the nonnegative energy density, the standard null -- and thus weak, strong and dominant energy conditions are always violated along the radial direction as $\rho_m+P_m^r<0$ for $r\geq r_0$; to make matters slightly better, Fig.~\ref{Fig8} indicates that in a sense such violation could be reduced for greater values of $\hat\omega$ in the domain $\hat\omega>-2/A$. Moreover, as shown in Fig.~\ref{Fig10} which fixes $\hat\omega=1$ and studies the influences of $A$ instead, we notice that even $\rho_m\geq0$ no longer holds throughout $r\geq r_0$ for a spatially quickly decaying ($A\gtrsim2$) scalar field.


%
%
%
%

Comparing Figs.~\ref{Fig8} $\sim$ \ref{Fig10} of Brans-Dicke gravity with Fig.~\ref{Fig3}, one could find that due to the presence of the potential hill $V(\phi)>0$ and the possibly antigravitational Gauss-Bonnet effect, LBD gravity could
``better'' protect the standard NEC when supporting wormholes.
The results in this section supplement the earlier investigations in Ref.\cite{Lobo Wormhole Brans-Dicke}.
Moreover, recall that in scalar-tensor theory with the total Lagrangian density  $\mathscr{L}=\frac{1}{16\pi G}\left[f(\phi)R-h(\phi)\cdot\nabla_\alpha\phi\nabla^\alpha\phi-2U(\phi) \right]+\mathscr{L}_m$ in the Jordan frame, we have the following conditions for the sake of ghost-freeness and quantum stability \cite{Positive Geff scalar-tensor}: the graviton is non-ghost if $f(\phi)>0$ (as mentioned before in Sec.~\ref{Subsec Generic conditions supporting static, spherically symmetric wormholes}), while the scalar field $\phi(x^\alpha)$ itself is non-ghost if
\begin{equation}
\frac{3}{2}\left(\frac{df(\phi)}{d\phi}\right)^2+f(\phi)h(\phi)>0\,.
\end{equation}
For Brans-Dicke gravity with $f(\phi)=\phi$ and $h(\phi)=\hat\omega/\phi$, it requires $\phi>0$ and $\hat\omega>-3/2$ to be totally ghost-free. Thus, the lessons from Figs.~\ref{Fig8} $\sim$ \ref{Fig10} are consistent with the argument of Ref.\cite{Positive Geff scalar-tensor} that in scalar-tensor gravity, there exists no  static, spherically symmetric wormholes  that are both ghost-free and obeying the standard NEC.

%



\section{Discussion and conclusions}

In Secs.~\ref{Sec Supporting wormholes in the LBD graity} and ~\ref{Sec An exact solution}, we have seen that $\rho_{\text{CP}}$ and $P_{\text{CP}}^r$ did not help in supporting Morris-Thorne wormholes; this is because $H_{\mu\nu}^{\text{(CP)}}$ identically vanishes for all spherically symmetric spacetimes (no matter static or dynamical). In fact, it can be directly verified that the spacetime parity will come into effect via nonzero $H_{\mu\nu}^{\text{(CP)}}$ in generic axially symmetric spacetimes, say the metric below that generalizes Morris-Thorne into rotating wormholes \cite{Rotating traversable wormholes}:
\begin{equation}\label{axisymmetric stationary rotating wormhole}
ds^2=-e^{2\Phi(r,\theta)} dt^2+  \left(1-\frac{b(r,\theta)}{r}\right)^{-1}dr^2
+r^2\Big[d\theta^2+\sin^2\theta\left(dt-\omega  d\varphi\right)^2\Big]\,,
\end{equation}
where, as in the Kerr or Papapetrou metric, $\omega=\omega(r,\theta)$ is the angular velocity $d\varphi/dt$ acquired by a test particle falling to the point $(r,\theta)$ from infinity.

Also, the wormhole geometry is not only related to the energy-momentum distribution of the physical matter through the gravitational field equation, but also to the propagation of the scalar field through the kinematical wave equation. Hence, in Secs.~\ref{Sec An exact solution} and \ref{Sec Wormholes in Brans-Dicke gravity}, for the sake of simplicity, we have chosen to ``recover'' the shape function $b(r)$ and thus the wormhole geometry from the kinematics of $\phi(r)$, i.e. Eqs.(\ref{kinematic wave equation for LBD}) and (\ref{kinematic wave equation for BD}), while the field equations were employed to analyze the energy conditions. When seeking for zero-tidal-force solutions with a vanishing redshift function $\Phi(r)=0$, this provides a simpler method than that in Ref.\cite{Lobo Wormhole Brans-Dicke} for Brans-Dicke gravity, or Ref.\cite{Lobo Wormhole hybrid metric-Palatini} for hybrid metric-Palatini $f(R)$ gravity which is equivalent to the mixture of GR and the $\hat\omega=-3/2$ Brans-Dicke gravity; they solve for $b(r)$ from the \textit{dynamical} wave equation (i.e. Klein-Gordon equation) rather than the \textit{kinematical} wave equation, and thus have to involve the trace of the matter tensor $T^{\text{(m)}}=-\rho_m+P_m^r+2P_m^T$ right from the beginning. However, when looking for more general solutions with $\Phi(r)\neq0$, one should still turn to the method in Refs.\cite{Lobo Wormhole Brans-Dicke} and \cite{Lobo Wormhole hybrid metric-Palatini}, as it becomes insufficient to determine the two Morris-Thorne functions $\{\Phi(r)\,,b(r)\}$ from a single kinematical wave equation.

To sum up, in this paper we have investigated the conditions to support traversable wormholes in LBD gravity. The flaring-out condition, which arises from  the wormholes' embedding geometry and thus applies to all metric gravities, requires the violation of the standard NEC in GR and the GNEC in modified gravities. Moreover, the breakdown of the null energy condition simultaneously violates the weak, strong and dominant energy conditions. With these considerations, we have derived the generalized energy conditions Eqs.(\ref{Null and Stong GECs}), (\ref{Weak GEC}), (\ref{LBD Null GEC}) and (\ref{LBD Weak GEC}) for LBD gravity in the form that explicitly contains the effective gravitational coupling strength $G_{\text{eff}}=\phi^{-1}$.  These energy conditions have been used to construct the conditions supporting Morris-Thorne-type wormholes, including the tensorial expressions Eqs.(\ref{Wormhole GEC Null}) and (\ref{Wormhole Weak GEC}), and their anisotropic-perfect-fluid forms in Sec.~\ref{Subsec Supporting conditions in anisotropic fluid scenario}. Moreover, in Sec.~\ref{Sec An exact solution} we have obtained an exact solution of the Morris-Thorne wormhole with a vanishing redshift function and the shape function Eq.(\ref{Exact solution br}), which is supplemented by the homogeneous scalar field $\phi(x^\alpha)=\phi(r)$ in Eq.(\ref{Homogeneous scalar field ansatz}) and the potential $V(\phi)=V_0\phi^2$. With the flexible parameters in Eq.(\ref{Exact solution br}) for $b(r)$ specified by  $\left\{\phi_0=r_0=V_0=1, A>0,  \omega_{\text L}<-2/A\right\}$, we have further confirmed the Lorentzian signature, the flaring-out condition, breakdown of the GNEC, and validity of the standard NEC. Finally, we also investigated zero-tidal-force wormholes in Brans-Dicke gravity, and have shown that the condition $\rho_m+P^r_m\geq0$ is not so well protected as in LBD gravity.


Note that natural existence of dark energy becomes effective  only at scales greater than 1Mpc \cite{Dark force scale}. Similarly in modified gravties, the higher-order terms or extra degrees of freedom are astrophysically recognizable only at galactic and cosmic levels. Hence, supporting wormholes by dark energy requires to mine and condense dark energy, while supporting wormholes by modified gravity requires unusual distributions of ordinary matter. For example, in LBD gravity, the joint effects of $H_{\mu\nu}^{\text{(CP)}}$, $H_{\mu\nu}^{\text{(GB)}}$ and $T_{\mu\nu}^{(\phi)}$ have to become dominant over the physical matter source $T_{\mu\nu}^{\text{(m)}}$, and the scalar field is preferred to be noncanonical. As a closing remark, we have to admit that wormholes in existent studies are mainly theoretical exercises and hypothetical objects, and there seems a long way ahead before wormholes can be artificially constructed and put to astronomical use.


\section*{Acknowledgement}

This work  was supported by NSERC grant 261429-2013.



\end{document}